\documentclass[twocolumn]{aastex63}
\usepackage{times}
\usepackage{amsmath}
\usepackage{textcomp}
\usepackage{amssymb}
\usepackage{natbib}
\usepackage{float}
\usepackage{color}
\usepackage{graphicx}

\newcommand{\gm}{$\gamma$}

\newcommand{\Ca}{Ca{\sevenrm II}}

 \font\sevenrm=cmr7 scaled 1000

\received{\today}
\revised{\today}
\accepted{\today}
\submitjournal{ApJ}

\shorttitle{Gamma-ray Emitting Misaligned AGN}
\shortauthors{Paliya et al.}

\begin{document}
\title{Radio Morphology of Gamma-ray Sources: Double-Lobed Radio Sources}

\correspondingauthor{Vaidehi S. Paliya}
\email{vaidehi.s.paliya@gmail.com}

\author[0000-0001-7774-5308]{Vaidehi S. Paliya}
\affiliation{Inter-University Centre for Astronomy and Astrophysics (IUCAA), SPPU Campus, Pune 411007, India}
\author[0000-0002-4464-8023]{D. J. Saikia}
\affiliation{Inter-University Centre for Astronomy and Astrophysics (IUCAA), SPPU Campus, Pune 411007, India}
\author[0000-0002-3433-4610]{Alberto Dom{\'{\i}}nguez}
\affiliation{IPARCOS and Department of EMFTEL, Universidad Complutense de Madrid, E-28040 Madrid, Spain}
\author[0000-0002-4464-8023]{C. S. Stalin}
\affiliation{Indian Institute of Astrophysics, Block II, Koramangala, Bengaluru 560034, Karnataka, India}

\begin{abstract}
The extragalactic \gm-ray sky is dominated by relativistic jets aligned to the observer's line of sight, i.e., blazars. A few of their misaligned counterparts, e.g., radio galaxies, are also detected with the Fermi-Large Area Telescope (LAT) albeit in a small number ($\sim$50), indicating the crucial role played by the jet viewing angle in detecting \gm-ray emission from jets. These \gm-ray emitting misaligned active galactic nuclei (AGN) provide us with a unique opportunity to understand the high-energy emission production mechanisms from a different viewpoint than the more common blazars. With this goal in mind, we have systematically studied the radio morphology of \gm-ray emitting sources present in the fourth data release of the fourth catalog of Fermi-LAT detected \gm-ray sources to identify misaligned AGN. By utilizing the high-resolution and sensitive MHz and GHz frequency observations delivered by the Very Large Array Sky Survey, Low-Frequency Array Two-metre Sky Survey, Faint Images of the Radio Sky at Twenty-Centimeters, and Rapid ASKAP Continuum Survey, here we present a catalog of 149 \gm-ray detected misaligned AGN, thus $\sim$tripling the number of known objects of this class. Our sample includes a variety of radio morphologies, e.g., edge-darkened and edge-brightened, hybrids, wide-angle-tailed, bent jets, and giants. Since the \gm-ray emission is thought to be highly sensitive to the jet viewing angle, such an enlarged sample of \gm-ray detected misaligned radio sources will permit us to explore the origin of high-energy emission in relativistic jets and radio lobes and study AGN unification, in general.

\end{abstract}


\section{Introduction}
Radio galaxies are thought to be the misaligned version of beamed active galactic nuclei (AGN), namely the blazars, which consist of flat-spectrum core-dominated quasars and BL Lac objects \citep[cf.][for a review]{1995PASP..107..803U}. A radio-loud quasar is considered to be a misaligned jetted AGN if the viewing angle of the jet $\theta_{\rm v}$ is $>1/\Gamma$, where $\Gamma$ is the bulk Lorentz factor of the jet. Because of the large viewing angle, the radiation emitted from either a radio galaxy or a misaligned quasar is less boosted compared to that observed from blazars. These sources at large viewing angles are mostly identified at radio wavelengths where the radiation is dominated primarily by the mildly relativistic outflows from the extended jets and hotspots or nearly isotropic emission originating from the radio lobes \citep[cf.][for a recent review]{2022JApA...43...97S}. Radio galaxies and quasars have been classified as Fanaroff-Riley Type I (FR I) and FR II sources based on their extended radio structure, with the former being less luminous with an edge-darkened structure and reasonably symmetric large-scale radio jets. The FR II sources are edge-brightened, usually with prominent hotspots along with asymmetric large-scale jets, and are found to be more luminous \citep[][]{1974MNRAS.167P..31F}. However, recent low-frequency observations taken with the Low-Frequency Array (LOFAR) have revealed that the radio luminosity may not reliably predict the morphology and classification of a radio galaxy \citep[][]{2019MNRAS.488.2701M}. Furthermore, hybrid morphology radio sources, i.e., those exhibiting an FR II radio structure on one side and an FR I jet on the other, have also been reported \citep[cf.][]{1996MNRAS.282..837S,2000A&A...363..507G}. More recently, there have been suggestions that the hybrid morphology sources are the result of orientation and are intrinsically FRII radio sources \citep[][]{2020MNRAS.491..803H}. The optical spectroscopic observations of radio galaxies have also led to their classification into high-excitation radio galaxies (HERGs) and low-excitation radio galaxies (LERGs), with the former having radiatively efficient accretion \citep[cf.][]{2012MNRAS.421.1569B,2014ARA&A..52..589H}. However, recent LOFAR observations have indicated a complex relationship between FR class and accretion mode in radio sources \citep[][]{2019MNRAS.488.2701M,2022MNRAS.511.3250M}.

In the high-energy \gm-ray band (0.1$-$300 GeV), the number of known radio galaxies is small ($\sim$50) compared to blazars which dominate the extragalactic \gm-ray sky \citep[e.g.,][]{2010ApJ...720..912A,2012ApJ...751L...3G,2020JHEAp..27...77C,2022ApJS..263...24A}. This can be understood due to the relatively large viewing angle causing de-boosting of the emitted radiation from the jet. Therefore, the detection of \gm-ray emission from radio galaxies may be indicative of a different emission region location and/or physical processes. Indeed, the extended \gm-ray emission originating from the lobes of the radio galaxy, Centaurus A, has been observed \citep[][]{2010Sci...328..725A}. \citet[][]{2022ApJ...931..138F} carried out a systematic study of \gm-ray detected radio galaxies present in the second data release of the fourth Fermi-Large Area Telescope (LAT) \gm-ray source catalog and determined that these objects may contribute up to $\sim$10\% of the extragalactic \gm-ray background. The \gm-ray emission in these objects has been typically explained with synchrotron self Compton radiative models \citep[e.g.,][]{2018ApJ...855...93F}. However, several complex models, e.g., spine-sheath structured jet, have also been put forward to explain the \gm-ray observations of radio galaxies \citep[][]{2010MNRAS.402.1649G,2011A&A...533A..72M}.

The identification of radio galaxies in the \gm-ray band has primarily been done by cross-matching the list of known radio galaxies with Fermi-LAT catalogs \citep[cf.][]{2010ApJ...720..912A}. A majority of these radio galaxies were found in dedicated observations and/or previous generation low-resolution surveys, e.g., NRAO VLA Sky Survey, that permitted us to identify mainly the large-scale radio structures extending up to Mpc scales \citep[e.g.,][]{1988AJ.....96...30C,1999MNRAS.309..100I,2023JApA...44...13D}. With the advent of sensitive, wide-field, high-resolution radio surveys, e.g., the Very Large Array Sky Survey (VLASS), it is now possible to identify radio galaxies with comparatively smaller linear sizes. Indeed, several \gm-ray sources that were not known as misaligned AGN have recently been found to exhibit radio structure typical of FR I/II radio galaxies \citep[cf.][]{2022MNRAS.513..886B,2023MNRAS.520L..33P}. Additionally, a new population of \gm-ray emitting FR 0 radio galaxies has also emerged in the \gm-ray sky \citep[e.g.,][]{2016MNRAS.457....2G,2021ApJ...918L..39P,2023ApJ...957...73P}. These observations suggest that there could be many more high-energy emitting misaligned AGN than currently known. They may have remained unresolved/partially resolved in previous surveys, possibly due to poorer spatial resolution. 

Enlarging the \gm-ray emitting radio galaxy population is crucial for sample studies, improving source classification of the Fermi-LAT catalogs, and constraining the contribution of radio galaxies to the extragalactic \gm-ray background \citep[cf.][]{2014ApJ...780..161D,2019ApJ...879...68S,2022ApJ...931..138F}. Furthermore, a larger sample also permits understanding the radiative processes responsible for the observed high-energy emission and studying the surrounding environment in which they grow and launch relativistic jets \citep[e.g.,][]{2023ApJS..265...60C}. With these objectives in mind, we utilized the latest multi-frequency radio surveys to study the morphological properties of all known \gm-ray sources. Here we report those \gm-ray sources that exhibit extended radio morphologies in at least one of the considered radio surveys. We also used the optical spectroscopic observations, when available, to identify \gm-ray detected misaligned AGN with high confidence. Section~\ref{sec2} briefly describes the sample selection and multi-wavelength radio catalogs whose data were used in this work. The results are presented and discussed in Sections~\ref{sec3} and \ref{sec4}, respectively. 
We summarize our findings in Section~\ref{sec6}. A flat cosmology with $H_0 = 70~{\rm km~s^{-1}~Mpc^{-1}}$ and $\Omega_{\rm M} = 0.3$ was adopted.

\section{Sample Selection and Multi-wavelength Catalog}\label{sec2}
\subsection{The Sample}
The fourth data release of the fourth Fermi-LAT \gm-ray source catalog (4FGL-DR4) contains 7194 astrophysical objects significantly detected at \gm-ray energies \citep[][]{2022ApJS..260...53A,2023arXiv230712546B}. Among them, we considered those 4069 \gm-ray sources whose multiwavelength counterparts have been reported to be AGN, e.g., blazars and radio galaxies. The radio morphologies of these objects were inspected using the data provided by the multi-frequency radio surveys described below. The counterpart coordinates of \gm-ray sources given in the 4FGL-DR4 catalog were used for this purpose. 

There are 2430 unidentified \gm-ray sources (unID) present in the 4FGL-DR4 catalog for which multi-wavelength counterparts remained unidentified. To determine the potential counterparts, we have utilized several multi-wavelength catalogs, e.g., Chandra Source Catalog, and considered only those sources located within the \gm-ray uncertainty region of unIDs that have been detected in the X-ray, radio, and optical/infrared (IR) bands. The details of the adopted steps are provided in Appendix (\ref{A:sec1}). We considered these 476 potential counterparts and studied their radio morphologies to identify \gm-ray emitting radio galaxies among them.

\subsection{Radio Catalogs}
We used the following radio surveys to study the morphologies of the sources under consideration:

Very Large Array Sky Survey (VLASS): It is an ongoing radio survey at 2$-$4 GHz covering the whole sky visible at VLA ($\delta>-40^{\circ}$). The angular resolution is $\sim$2.5 arcseconds and rms sensitivity for a single epoch is 120 $\mu$Jy/beam. The observations are being carried out in three epochs to study transient astrophysical objects. Further details about the survey can be found in \citet[][]{2020PASP..132c5001L}. We downloaded the $5\times 5$ arcmin$^{2}$ cutout images of 3676 sources. The images were downloaded from the Canadian Initiative for Radio Astronomy Data Analysis website\footnote{\url{http://cutouts.cirada.ca}}.

Faint Images of the Radio Sky at Twenty-Centimeters (FIRST): It is the previous generation survey conducted at 1.4 GHz with a typical rms sensitivity of 0.15 mJy/beam, and a resolution of $\sim$5 arcseconds \citep[][]{2015ApJ...801...26H}. We downloaded the $5\times 5$ arcmin$^{2}$ cutout images of 1300 sources from the FIRST cutout server \footnote{\url{https://third.ucllnl.org/cgi-bin/firstcutout}}.

Low-Frequency Array: The second data release of the LOFAR Two-metre Sky Survey (LOTSS-DR2) covers areas of 4178 and 1457 square degrees centered at (12$^{h}$ 45$^{m}$, +44$^{\circ}$ 30$^{\prime}$) and (1$^{h}$ 00$^{m}$ +28$^{\circ}$ 00$^{\prime}$). In the 120$-$168 MHz frequency band, a resolution of 6 arcseconds and a median rms sensitivity of 83 $\mu$Jy/beam have been achieved \citep[][]{2022A&A...659A...1S}. We retrieved $30\times 30$ arcmin$^{2}$ cutout images of 770 objects from the LOFAR data server\footnote{\url{https://lofar-surveys.org/dr2\_release.html}}.

Rapid ASKAP Continuum Survey (RACS): This ongoing survey utilizes the full 36-dish Australian Square Kilometre Array Pathfinder (ASKAP) telescope. A new catalog of sources detected at 1367.5 MHz (RACS-mid) was released, covering the sky south of declination +49$^{\circ}$ \citep[][]{2024PASA...41....3D}. The median angular resolution of the survey is 11$^{\prime\prime}$.2$\times$9$^{\prime\prime}$.3 though it varied to maximize sky coverage and sensitivity across the covered area of $\sim$36200 deg$^{2}$. We extracted $30\times 30$ arcmin$^{2}$ cutout images of 1267 objects from this catalog \footnote{\url{https://research.csiro.au/casda/}}. 

The choice of different cutout sizes was based on the resolution and the operating frequency of the adopted surveys. For example, VLASS has the highest resolution among all the surveys, hence, the large-scale diffuse emission is expected to be completely resolved out, thus undetected. Also the largest angular size of the known double-lobed radio sources in the VLASS is only $\sim$3 arcmin \citep{2023ApJS..267...37G}. Therefore, small-sized cutouts were downloaded. On the other hand, low-frequency surveys, such as RACS and LOFAR, are expected to detect diffuse emission mainly originating from the large-scale lobes, so larger cutouts were retrieved from their databases.

\section{Results}\label{sec3}
We retrieved and visually inspected the radio cutout images from the above-mentioned surveys. We started with the VLASS data and segregated the sources into two lists: one for sources whose VLASS images revealed double-lobed morphology and the other for remaining objects. The radio morphologies of sources present in the second list were then inspected using the FIRST images when available, and the source lists were updated. This procedure was then repeated for the LOFAR and RACS surveys, respectively.

Since the VLASS quick-look images can be affected by artifacts, it is possible that an extended radio structure may not be clearly visible due to image distortion and/or the presence of bright streaks. Alternatively, a false detection can appear like an extended radio jet\footnote{\url{https://science.nrao.edu/vlass/data-access/vlass-epoch-1-quick-look-users-guide}}. Also, this high-frequency, high-resolution survey may resolve out diffuse extended emission. To take into account these issues and for consistency check, the radio images taken in other surveys, when available, were inspected for all \gm-ray sources whose radio morphology was found to exhibit double-lobed structure in at least one of the considered surveys. Moreover, since VLASS quick-look images are available for several epochs, we ensured that an extended radio structure is detected in all of them. 

In addition to the double-lobed morphology, which can be affected by projection effects, we adopted the following three techniques to identify the best \gm-ray emitting misaligned AGN candidates: (i) visual inspection of the optical spectrum, (ii) calculation of the core dominance, and (iii) overall radio spectral index.

According to the AGN unification scheme, the strength of the broad emission lines decreases as the viewing angle increases and the radiation from the host galaxy stellar population starts dominating. Therefore, an optical spectrum consisting primarily of narrow emission lines and/or galaxy spectral features, e.g., \Ca~H/K break, indicates a large viewing angle. With this idea in mind, we also collected the available optical spectra from the literature for all sources exhibiting double-lobed radio morphologies.

The core dominance is considered a good proxy for the orientation of the beamed emission since the core emission is Doppler boosted, whereas the lobe emission can be assumed reasonably isotropic \citep[cf.][]{1997MNRAS.284..541M}. This parameter was derived from the following relation:

\begin{equation}
    C_{\rm D}=\log\left(\frac{F_{\rm core}}{F_{\rm ext}}(1+z)^{\alpha_{\rm core}-\alpha_{\rm ext}} \right),
\end{equation}

The parameters $\alpha_{\rm core}$ and $\alpha_{\rm ext}$ are spectral indices 
{($F(\nu) \propto \nu^\alpha$)}
of the core and extended emission, respectively, and were taken as $\alpha_{\rm core}=0$ and $\alpha_{\rm ext}=-0.8$. The parameters $F_{\rm core}$ and $F_{\rm ext}$ ($=F_{\rm total}-F_{\rm core}$) refer to the core and extended flux densities, respectively, and were estimated at rest-frame 3 GHz assuming the spectral indices mentioned above. For sources without redshift information, we derived the core dominance as $C_{\rm D}=\log\frac{F_{\rm core}}{F_{\rm ext}}$, i.e., at the observed 3 GHz frequency. For blazars, $C_{\rm D}$ is reported to be large \citep[$>$1;][]{2001MNRAS.326.1455M,2015MNRAS.451.4193C}. On the other hand, using Very Long Baseline Interferometry (VLBI) observations, \citet[][]{2019A&A...627A.148A} found the core dominance of the Fermi-LAT detected radio galaxies to be $<1$. 

\begin{figure*}
\hbox{\hspace{1.cm}
    \includegraphics[trim={0 0 0 0},clip, scale=0.75]{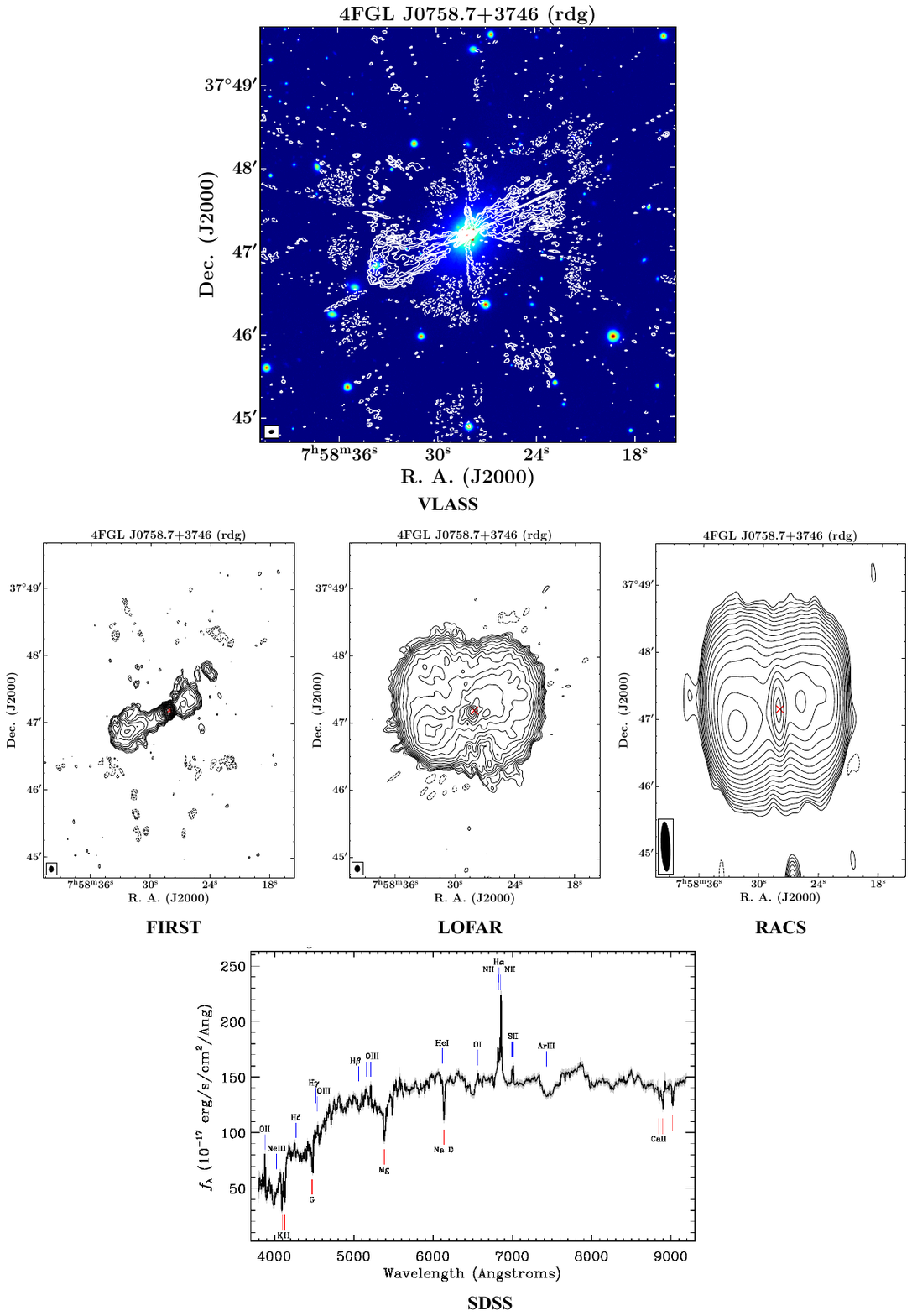}
    }
\caption{The VLASS contours overplotted on PanSTARRS $i$-band images of one of the \gm-ray emitting AGN exhibiting double-lobed radio structure (top panel). Along with the 4FGL name, we also mention the classification reported in the 4FGL-DR4 catalog \citep[][]{2022ApJS..260...53A}. The radio morphologies observed by other surveys, when available, are plotted in the middle panel. In all plots, the contour levels start at 3$\times$rms$\times$($-$2, $-\sqrt{2}$, $-$1, 1) and increases in multiple of $\sqrt{2}$. The bottom panel shows the optical spectrum of the source when available. By default, the north is up and east to the left. For DECAM optical images, we have added a compass to guide the eye. The references for shown optical spectra are provided in Table~\ref{tab:basic_info}. The complete figure set is available on ZENODO (doi: 10.5281/zenodo.13908693).}\label{fig:vlass1}
\end{figure*}

The overall radio spectral shape of an AGN can also give hints about the jet misalignment. Sources viewed close to the line of sight, i.e., blazars, usually exhibit a flat radio spectrum throughout MHz-to-GHz frequencies \citep[spectral index $\alpha > -0.5$,][]{2002AJ....124.2364I,2007ApJS..171...61H,2014ApJS..213....3M}. On the other hand, radio galaxies often exhibit steep radio spectra. Depending on the degree of core dominance at different frequencies, sources may also have a steep spectrum at low frequencies due to the extended emission and a flat spectrum at high frequencies where the core dominates. We used the SPECFIND (v3.0) catalog to determine the radio spectral indices of the sources under consideration \citep[][]{2021A&A...655A..17S}. For 8 objects not present in this catalog, we determined their spectral indices by taking the integrated flux densities from the NRAO VLA Sky Survey (1.4 GHz), TIFR GMRT Sky Survey (150 MHz), LOFAR (144 MHz), GaLactic and Extragalactic All-sky MWA (GLEAM) survey (200 MHz), and the Sydney University Molonglo Sky Survey (843 MHz) depending on the presence of objects in the footprints of these surveys.

The visual inspection of all cutout images led to the identification of 219 sources where double-lobed radio emission was noticed in at least one of the above-mentioned radio surveys. We then devised a strategy that a source is considered as a promising \gm-ray emitting misaligned AGN only if at least two of the adopted three observables, i.e., optical spectrum, core dominance, and radio spectral index, resemble known misaligned radio sources. This exercise led to the final sample of 149 \gm-ray emitting objects. 

We were able to retrieve the optical spectra of 103 sources, including two of them from our own observations taken at Telescopio Nazionale Galileo (Nuria Álvarez Crespo et al., in preparation). The remaining 46 objects either do not have optical spectroscopic observations or the spectral plots are not published. There are 86 sources whose optical spectra are dominated by host galaxy absorption features, whereas 17 sources exhibit broad emission lines in their optical spectra. Considering the radio spectra, 116 sources have $\alpha<-0.5$, i.e., a steep radio spectrum. Among the remaining 33 sources, 8 have a radio spectral index close to the steep/flat spectral shape boundary, i.e., $-0.5<\alpha<-0.45$. The logarithmic core dominance is found to be less than 0.5 for 147, and only two sources have core-dominated emissions. Among these 147 objects, 106 have lobe-dominated emission, i.e., $C_{\rm D}<0$. Finally, we also classified sources as FR I and IIs based on their observed morphologies. There are 64 and 71 objects exhibiting FR I and II radio structures, respectively. We have also found 14 \gm-ray emitters that show a hybrid radio morphology, i.e., FR I type diffuse emission on one side and edge-brightened FR II lobes on the opposite side of the core. We show one of their radio images and optical spectrum in Figure~\ref{fig:vlass1} and list all sources in Table~\ref{tab:basic_info}. The contours from the radio survey data in which the double-lobed morphology was the best seen were overplotted on the Panoramic Survey Telescope and Rapid Response System \citep[PanSTARRS;][]{2016arXiv161205560C} or Dark Energy Survey \citep[DES;][]{2021ApJS..255...20A} $i$-band images. For completeness, the radio contours using other survey data, when available, were also plotted. The plots for the remaining objects can be found on Zenodo\footnote{\url{https://doi.org/10.5281/zenodo.13908693}}.

We estimated the total flux density using the survey data with the largest beam size, i.e., lowest resolution, to ensure that no diffuse extended flux density is missed out. The core flux density, on the other hand, was derived using the highest resolution survey data to ensure that the core emission is well disentangled from the extended lobes. The largest angular size was measured using the survey data in which the object under consideration was best resolved while also considering that low-resolution surveys are likely to pick the low surface brightness extended flux density. For sources exhibiting FR II morphology, we estimated the largest angular size from the brightest pixel in the hotspot. For FR Is, on the other hand, it was computed from the 3$\sigma$ outermost flux density contour level. For this purpose, the rms noise value ($\sigma$) was estimated from a nearby source-free region. Moreover, the largest angular size was computed by measuring the lengths along the jet/lobe from the core for both sides of the jet/lobe, which were then added. We caution that the true source length could be different due to projection effects, so our measurements are purely observational in nature. The core flux density was calculated by considering a circular region centered at the radio position and having the size radius as the resolution of the adopted survey. As the resolution of the RACS data significantly varies as a function of the source location, there may not be an adequate number of resolution elements across the source axis to clearly resolve the radio core. In such cases, we relied on the beam size at the location of the optical object to estimate the core flux density. In these cases, the core flux density would be an upper limit. On the other hand, the total flux density was estimated by integrating all the flux density within the outermost 3$\sigma$ contour level. These parameters are provided in Table~\ref{tab:basic_info}.

\section{Discussion}\label{sec4}
The 4FGL-DR4 catalog reports 53 \gm-ray emitting radio galaxies along with 2 steep-spectrum radio quasars. In this work, we have found 149 \gm-ray sources that exhibit double-lobed radio morphology and other observational properties, such as radio spectral shape, resembling misaligned radio sources. Out of 53 known \gm-ray emitting radio galaxies, 41 are present in our sample. We briefly discuss the remaining 12 objects in Appendix~\ref{app:missing}.

\begin{figure*}
\gridline{\fig{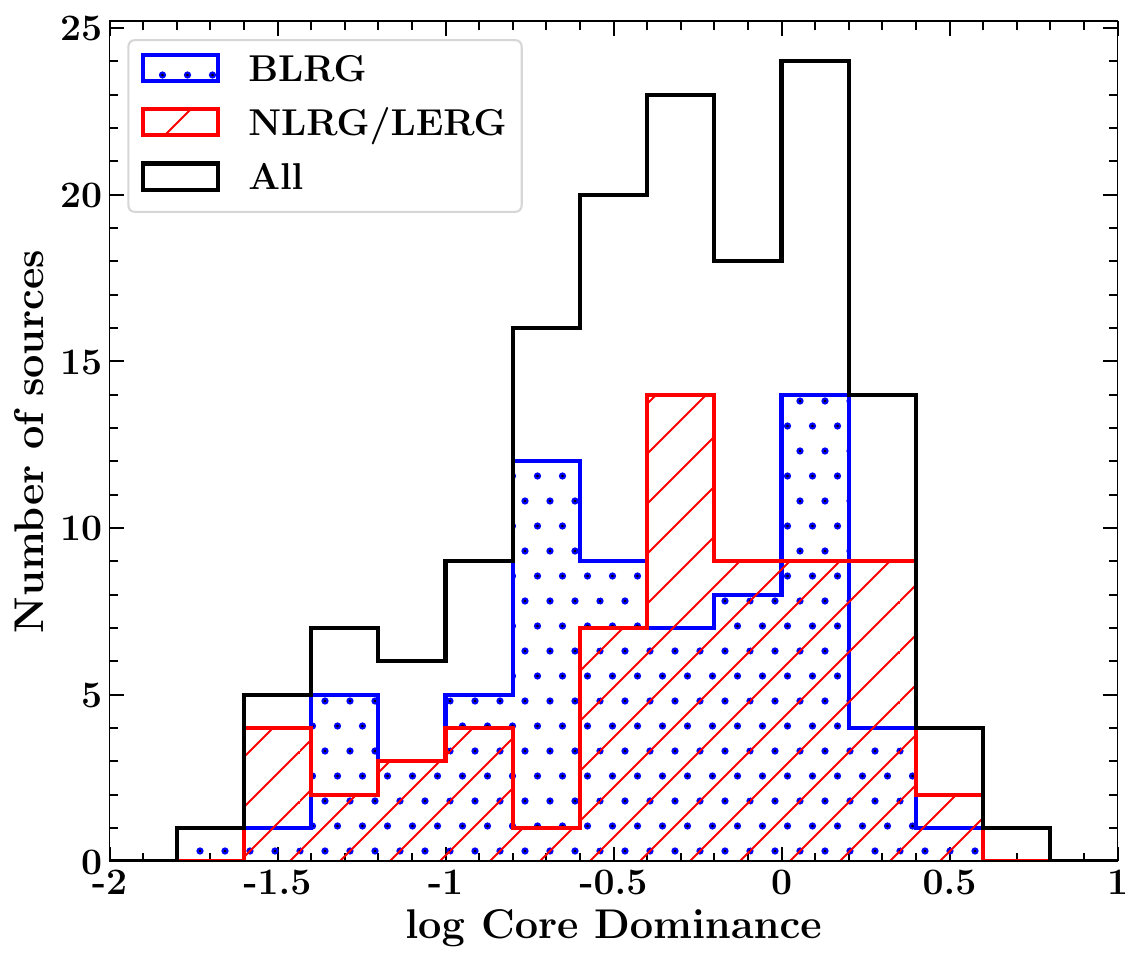}{0.32\textwidth}{(a)}
          \fig{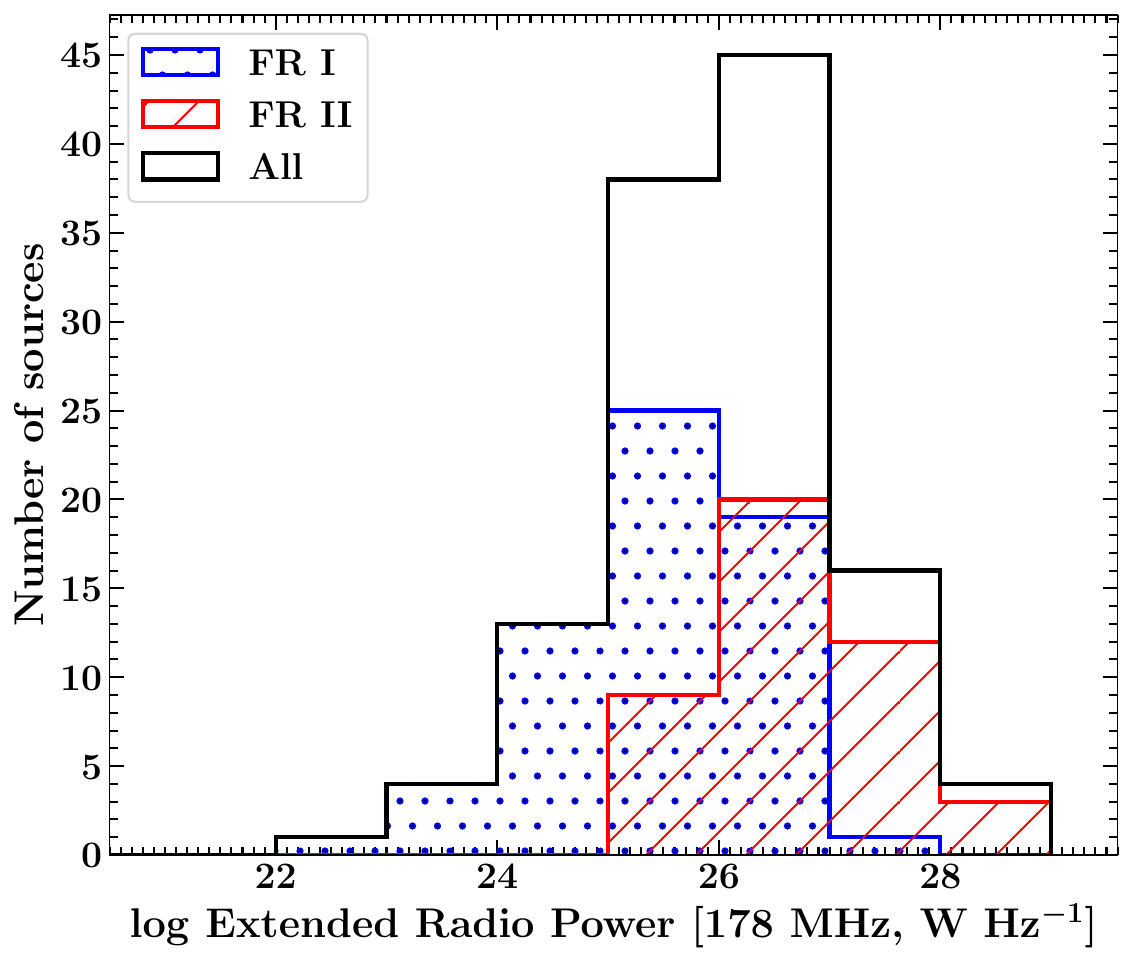}{0.32\textwidth}{(b)}
          \fig{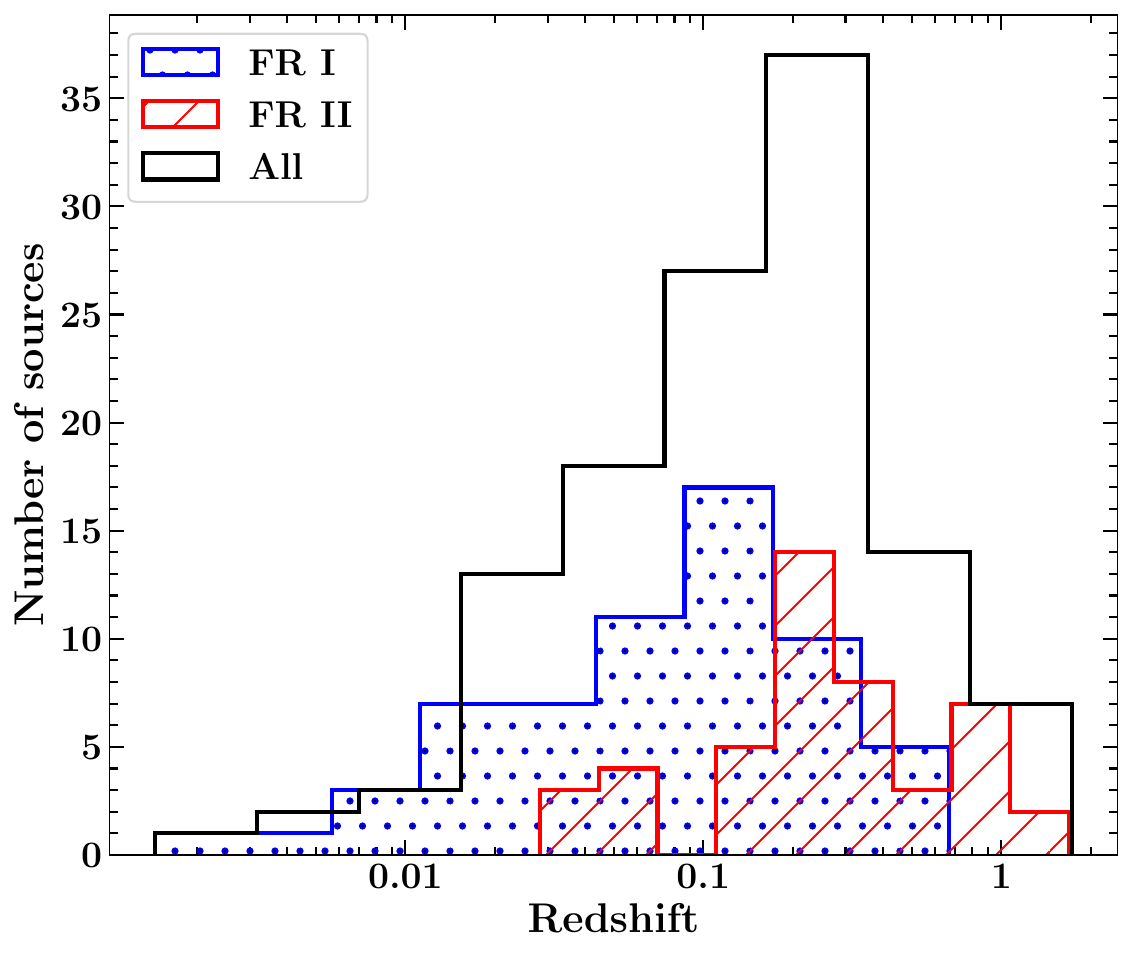}{0.32\textwidth}{(c)}
          }
\gridline{\fig{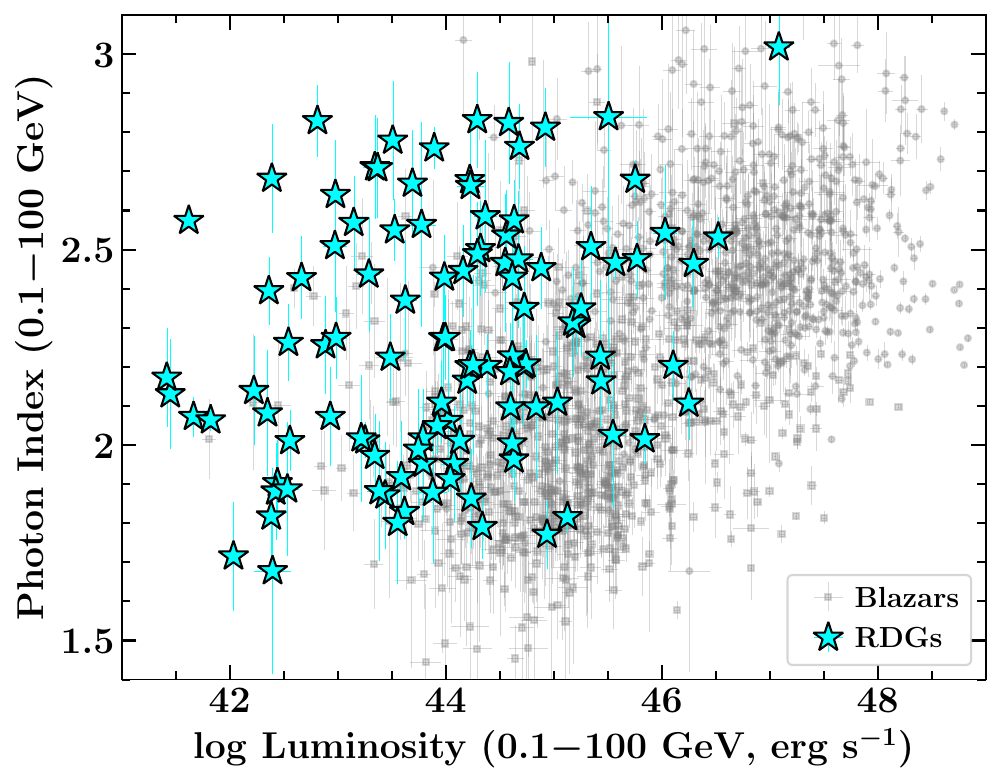}{0.33\textwidth}{(d)}
          \fig{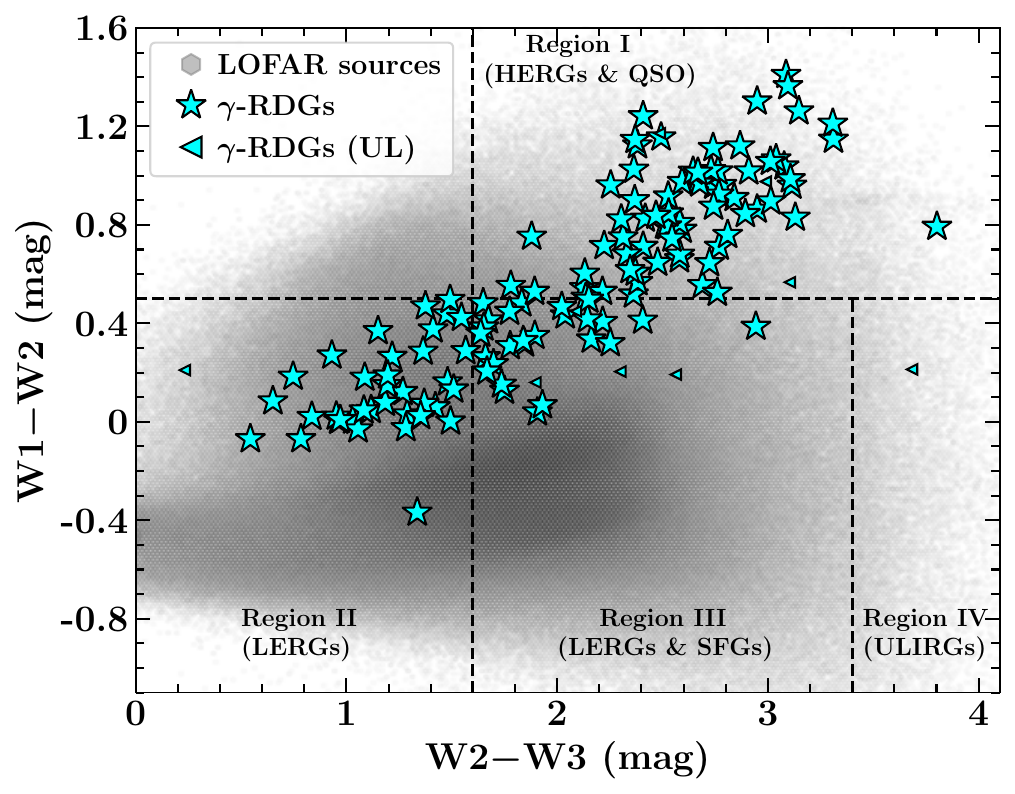}{0.33\textwidth}{(e)}
          \fig{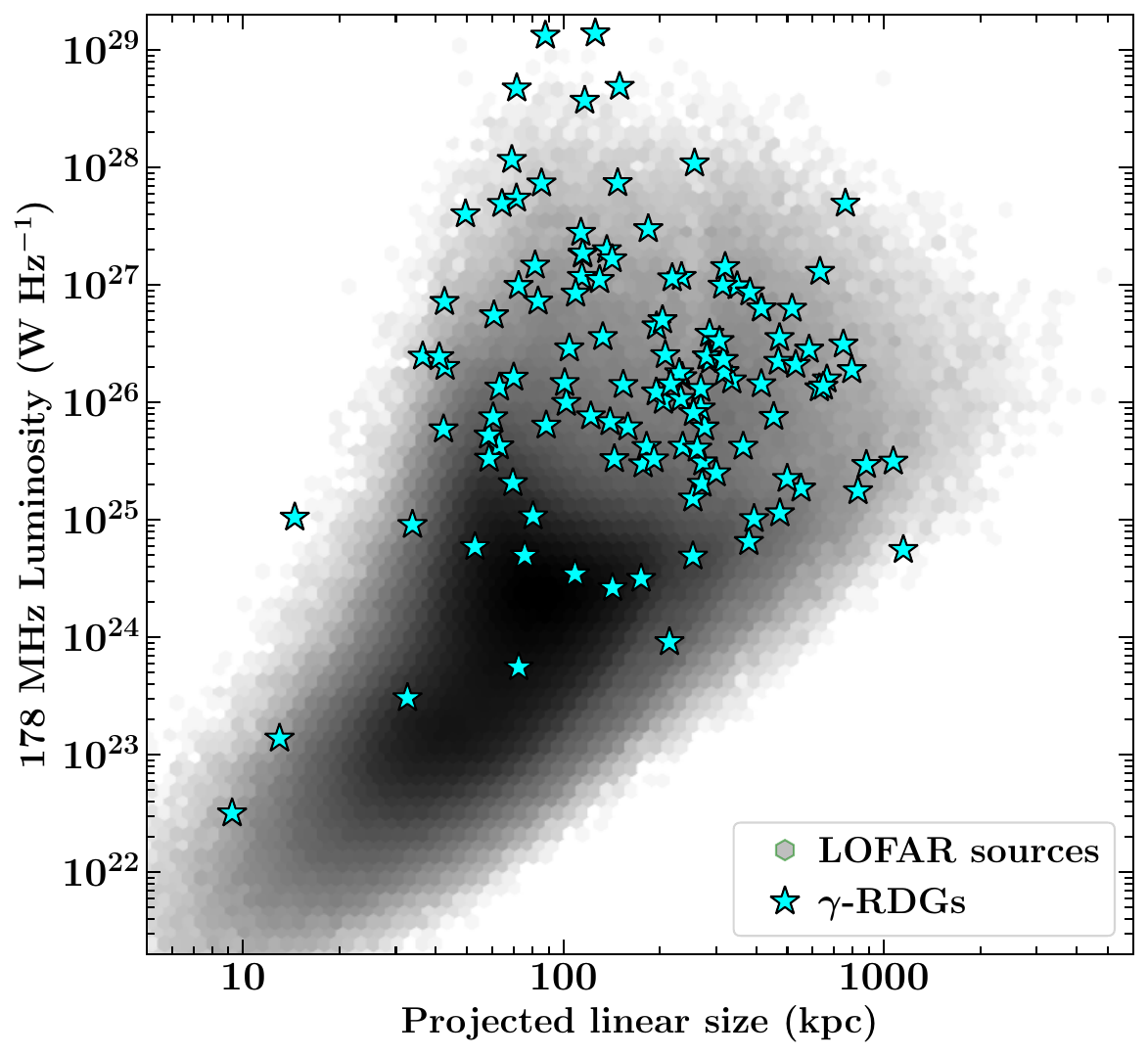}{0.28\textwidth}{(f)}
          }
\caption{The histograms of the core dominance (panel (a)), radio power (panel (b)), and redshift (panel (c)) for 149 misaligned AGN found in this work. In the top left panel, the blue-dotted and red-hatched histograms refer to objects whose optical spectra exhibit broad emission lines and host galaxy absorption features or narrow emission lines, respectively. In the top middle and right panels, these correspond to objects showing FR I and II morphologies, respectively, as found in this work. The panel (d) shows the \gm-ray luminosity versus photon index plot for Fermi-LAT detected jetted sources. The panels (e) and (f) correspond to the WISE color-color diagram and total radio power versus projected jet length plot, respectively. In all the panels, the \gm-ray detected misaligned AGN are highlighted with cyan stars. The triangles shown in panel (e) refer to upper limit in the $W3$-band magnitude. The grey data points belong to the Fermi-LAT blazars (panel (d)), and LOFAR detected sources (panels (e) and (f)).}
\label{fig:cd}
\end{figure*}

A $\sim$three-fold increase in the number of \gm-ray emitting misaligned AGN is likely due to much better sensitivity and resolution of the latest radio surveys that enabled us to identify faint extended radio structures. Furthermore, prior works have reported the detection of more \gm-ray detected FR I sources than FR IIs \citep[][]{2022ApJS..263...24A}. This could be due to the closer proximity of the former and/or different beaming factors \citep[cf.][]{2010ApJ...720..912A,2022ApJ...931..138F}. Since FR IIs are expected to have radiatively efficient accretion, their \gm-ray emission might be produced due to the external Compton process, which is more sensitive to the beaming effect compared to synchrotron self Compton mechanism \citep[e.g.,][]{1994ApJ...421..153S,1995ApJ...446L..63D}. However, optical spectroscopic observations have found several radiatively inefficient LERGs associated with FR II radio morphologies \citep[e.g.,][]{2009ANA...495.1033B}, thereby raising questions on their \gm-ray production mechanism. Indeed, we have found a comparable number of \gm-ray emitting FR I and II radio sources in this work. The identification of a large number of \gm-ray emitting FR II sources suggests that the \gm-ray production mechanisms could be more complex than that explained using a standard one-zone leptonic radiative model \citep[cf.][]{2010MNRAS.402.1649G,2011A&A...533A..72M}. 

The core dominance is a good proxy for the orientation with respect to our line of sight \citep[e.g.,][]{1997MNRAS.284..541M}. This parameter is found to be larger for the Fermi-LAT detected misaligned AGN than their \gm-ray undetected counterparts \citep[][]{2010ApJ...720..912A}. The median logarithmic core dominance for our misaligned AGN sample is $-$0.302 ranging from $-$2.245 to 0.614 (Figure~\ref{fig:cd}, panel (a)). Dividing the sources based on their optical spectral properties, we found that objects having LERG-type spectra have smaller core dominance (median $C_{\rm D}=-0.245$) than those exhibiting broad emission lines (median $C_{\rm D}=-0.04$, Figure~\ref{fig:cd}, panel (a), where all broad emission-line objects have been listed as BLRGs). Though the derived values are comparatively higher than that noticed for \gm-ray undetected radio galaxies, the observed trend is similar to that reported in previous studies \citep[cf.][]{1997MNRAS.284..541M}. Furthermore, dividing the sources based on their observed morphologies, we found that FR I sources tend to be more core-dominated (median $C_{\rm D}=-0.239$) compared to FR IIs (median $C_{\rm D}=-0.461$). This is consistent with the known trend for the degree of core dominance to decrease with an increase in total radio luminosity. For example, in a study of the low-luminosity B2 radio galaxies, the majority of which were FR Is, along with the more luminous 3CR radio galaxies, most of which belong to the FR II category, core dominance was found to systematically decrease over about five orders of magnitude of radio luminosity \citep[e.g.][and references therein]{1990A&A...227..351D}.

The panel (b) of Figure~\ref{fig:cd} shows the distribution of the radio power measured at rest-frame 178 MHz frequency. For this purpose, we extrapolated the observed extended flux density values of 123 sources with known redshifts to the rest-frame 178 MHz assuming a spectral index $\alpha=-0.8$. The logarithmic median for the whole sample is 26.07 (in W Hz$^{-1}$). Dividing the sources based on their observed FR I and II morphologies, the median values are 25.51 and 26.83, respectively. These results are aligned with the traditional threshold of 10$^{26}$ W Hz$^{-1}$ with FR Is having lower luminosities. However, considering individual sources, exceptions are found in both classes, i.e., low-power FR IIs and high-power FR Is \citep[see also][]{2019MNRAS.488.2701M}.

The redshift distributions of FR I and II sources are shown in panel (c) of Figure~\ref{fig:cd}, and the estimated median values are 0.09 and 0.24, respectively. As also found in earlier works, FR IIs are located farther than FR Is. Improved sensitivity must have contributed to the \gm-ray detection of many FRII sources, which are, on average, located at higher redshifts. Another possibility could be the harder \gm-ray spectra of FR Is making them easier to detect with Fermi-LAT \citep[e.g.,][]{2010ApJ...720..912A}. With the addition of more data, hence deeper sensitivity, more soft spectrum FR IIs have now been detected. This is evident in panel (d) of Figure~\ref{fig:cd} where we show a plot of \gm-ray luminosity versus photon index considering blazars present in the 4FGL-DR4 catalog. A major fraction of the misaligned AGN tend to occupy a region of soft \gm-ray spectra (photon index $>$2) and low luminosity. These results are consistent with recent works \citep[see, e.g.,][]{2022MNRAS.513..886B}.

The panel (e) of Figure~\ref{fig:cd} shows the Wide-field Infrared Survey Explorer (WISE) color-color diagram. Earlier works have shown that WISE colors can be used to distinguish accretion modes in jetted AGN \citep[][]{2014MNRAS.438.1149G}. We have used the WISE color distinction reported in \citet[][]{2020A&A...642A.153D} and plotted 149 \gm-ray detected misaligned AGN. For a comparison, we have also shown the WISE colors of LOFAR detected sources \citep[][]{2023A&A...678A.151H}. In this diagram, the sample of \gm-ray sources mainly occupies the regions of HERGs and quasars (region I) and that of LERGs (region II). A few sources also lie in the region of LERGs and star-forming galaxies (SFGs).

The power/linear-size plane \citep[$P-D$ diagram;][]{1982IAUS...97...21B,1997MNRAS.292..723K} for LOFAR detected objects is shown in panel (f) of Figure~\ref{fig:cd} using the data published by \citet[][]{2023A&A...678A.151H}. To include the sample of the \gm-ray detected misaligned AGN, we extrapolated their observed total flux densities to that at rest-frame 178 MHz assuming a spectral slope of $-$0.8.  In this diagram, the \gm-ray sources tend to lie in a region of higher jet luminosity and a projected linear size spread around a few hundreds of kpc. The $P-D$ diagram has been extensively used to explore the evolution of radio-loud AGN; however, the location of \gm-ray detected sources on this plane should be studied with the caution that they are expected to have stronger Doppler boosting effects \citep[cf.][]{2015ApJ...810L...9L,2019A&A...622A..12H}.

 \startlongtable
\begin{deluxetable*}{ccccccccccccc}
\tabletypesize{\scriptsize}
\tablecaption{The list of \gm-ray emitting misaligned jetted AGN.\label{tab:basic_info}}
\tablewidth{0pt}
\tablehead{
\colhead{4FGL Name} & \colhead{RA} & \colhead{Dec} & \colhead{$z$} & \colhead{$\Gamma_{\gamma}$} & \colhead{$\alpha$} & \colhead{LAS} & \colhead{$F_{\rm core}$} & \colhead{$F_{\rm total}$} & \colhead{$C_{\rm D}$} & \colhead{Morph.} & \colhead{Survey} & \colhead{Ref.}\\
\colhead{[1]} & \colhead{[2]} & \colhead{[3]} & \colhead{[4]} & \colhead{[5]} & \colhead{[6]} & \colhead{[7]} & \colhead{[8]} & \colhead{[9]} & \colhead{[10]} & \colhead{[11]} & \colhead{12} & \colhead{13}}
\startdata
  J0001.4$-$0010  &0.33954 &$-$0.19441   &0.461 & $2.11\pm0.18$ &$-$0.57 &70.5    &$32.29  \pm0.26$ &$113.49   \pm0.78 $  &0.006  &FRI    &R  V  R & A20\\
  J0009.7$-$3217  &2.39816 &$-$32.27691  &0.025 & $2.26\pm0.10$ &$-$0.45 &211.2   &$210.77 \pm0.46$ &$445.29   \pm1.66 $  &0.237  &FRI    &R  V  R &  J09\\
  J0013.6+4051    &3.37970 &40.86031     &0.255 & $2.28\pm0.14$ &$-$0.53 &34.1    &$539.50 \pm0.61$ &$1600.79  \pm2.44 $  &0.060  &FRI$-$II &V  V  R &  S93\\
  J0014.2+0854    &3.58224 &8.90056      &0.163 & $2.50\pm0.12$ &$-$0.60  &118.9   &$74.54  \pm0.22$ &$362.44   \pm1.15 $  &$-$0.260 &FRI    &R  V  R &  A20\\
  J0018.8+2611    &4.91575 &26.04788     &0.280 & $2.81\pm0.10$ &$-$0.52 &54.6    &$272.48 \pm0.63$ &$756.53   \pm1.21 $  &0.111  &FRII   &V  V  R &  Z12\\
  J0037.9+2612    &9.32983 &26.22008     &0.148 & $2.45\pm0.12$ &$-$0.37 &411.1   &$79.01  \pm0.29$ &$556.61   \pm2.22 $  &0.322  &FRI    &L  V  L &  A20\\
  J0038.7$-$0204  &9.58553 &$-$2.12792   &0.220 & $2.76\pm0.11$ &$-$0.74 &19.9    &$509.48 \pm0.31$ &$6297.47  \pm4.00 $  &$-$0.711 &FRII   &V  V  R &  A20\\
  J0044.9+4553    &11.25014 &45.92090    &0.179 & $2.49\pm0.13$ &$-$0.59 &14.0    &$190.74 \pm0.38$ &$377.22   \pm0.74 $  &0.342  &FRII   &V  V  R &  Z12\\
  J0057.7+3023    &14.45368 &30.35244    &0.016 & $2.39\pm0.09$ &$-$0.30  &3410.1  &$528.56 \pm0.63$ &$10495.50 \pm10.52$  &$-$0.215 &FRI    &L  V  L &  J00\\
  J0112.8$-$0253  &18.20390 &$-$2.91769  &0.188 & $2.06\pm0.18$ &$-$0.73 &86.8    &$14.54  \pm0.36$ &$102.24   \pm0.98 $  &$-$0.445 &FRI    &R  V  R &  A20\\
  J0119.6+4158    &20.01148 &42.00386    &0.109 & $2.37\pm0.15$ &$-$0.60  &117.71  &$29.25  \pm0.26$ &$246.29   \pm1.20 $  &$-$0.560 &FRI    &L  V  R &  D20\\
  J0126.5$-$1553  &21.78538 &$-$15.93171 &0.988 & $2.54\pm0.17$ &$-$0.58 &93.7    &$26.66  \pm0.25$ &$107.36   \pm0.78 $  &0.033  &FRII   &R  V  R &  F22\\
  J0153.4+7114    &28.35771 &71.25179    &0.022 & $1.90\pm0.11$ &$-$0.43 &118.4   &$367.60 \pm0.46$ &$543.03   \pm2.54 $  &0.329  &FRI    &V  V  V &  M96\\
  J0204.3+2417    &31.08977 &24.29741    &0.210 & $1.95\pm0.15$ &$-$0.61 &68.0    &$81.46  \pm0.38$ &$213.27   \pm0.78 $  &0.132  &FRI    &L  V  R &  D20\\
  J0222.7+5016    &35.76560 &50.27069   &$-$999 & $2.17\pm0.12$ &$-$0.80  &17.6    &$5.41   \pm0.32$ &$25.17    \pm0.75 $  &$-$0.563 &FRII   &V  V  V &  \\
  J0237.7+0206    &39.42408 &1.97531     &0.021 & $2.14\pm0.14$ &0.00   &487.8   &$23.07  \pm0.24$ &$164.32   \pm2.06 $  &$-$0.505 &FRI    &R  V  R &  Z12\\
  J0305.1$-$1608  &46.31278 &$-$16.13796 &0.311 & $1.77\pm0.09$ &$-$0.77 &24.9    &$29.62  \pm0.31$ &$902.12   \pm1.55 $  &$-$1.100 &FRII   &V  V  R &  P17\\
  J0308.4+0407    &47.10924 &4.11091     &0.029 & $2.01\pm0.05$ &$-$0.59 &208.2   &$873.09 \pm0.85$ &$7510.69  \pm4.33 $  &$-$0.596 &FRI    &R  V  R &  B09\\
  J0309.9$-$6941  &47.20726 &$-$69.77450 &$-$999& $2.46\pm0.15$  &$-$0.42 &18.2    &$7.03   \pm0.09$ &$31.97    \pm0.28 $  &$-$0.275 &FRII   &R  R  R &  \\
  J0309.4$-$4000  &47.30425 &$-$40.03077 &0.193 & $2.01\pm0.13$ &$-$0.77 &145.0   &$124.20 \pm0.31$ &$352.28   \pm1.57 $  &0.072  &FRI    &R  R  R &  P21\\
  J0316.8+4120    &49.17907 &41.32489    &0.018 & $1.88\pm0.12$ &$-$0.22 &370.0   &$125.51 \pm1.19$ &$627.98   \pm2.80 $  &$-$0.321 &FRI    &R  V  R &  A20\\
  J0320.6+2728    &50.16735 &27.46344    &0.060 & $2.44\pm0.15$ &$-$0.34 &473.0   &$8.40   \pm0.25$ &$396.69   \pm1.81 $  &$-$1.370 &FRI    &R  V  R &  H12\\
  J0321.3$-$1612  &50.29293 &$-$16.21122 &0.459 & $2.10\pm0.18$ &$-$0.70  &87.9    &$1.60   \pm0.39$ &$114.52   \pm0.93 $  &$-$1.443 &FRI    &R  V  R &  F22\\
  J0322.6$-$3712e &50.67412 &$-$37.20820 &0.005 & $2.07\pm0.05$ &$-$0.77 &106.8   &$63.07  \pm0.41$ &$348.52   \pm1.95 $  &$-$0.379 &FRI    &R  V  R &  L17\\
  J0334.3+3920    &53.57672 &39.35677    &0.020 & $1.82\pm0.10$ &$-$0.55 &180.6   &$276.33 \pm0.42$ &$988.40   \pm1.80 $  &$-$0.129 &FRI    &R  V  R &  Z12\\
  J0418.2+3807    &64.58865 &38.02661    &0.048 & $2.76\pm0.05$ &$-$0.74 &204.3   &$1456.67\pm1.71$ &$14984.60 \pm15.69$  &$-$0.677 &FRII   &R  V  R &  B09\\
  J0431.0+3529c   &67.63673 &35.44046   &$-$999 & $2.64\pm0.14$ &$-$0.73 &56.5    &$1.91   \pm0.26$ &$66.82    \pm0.87 $  &$-$1.256 &FRII   &V  V  R &  \\
  J0515.5$-$0125  &78.90076 &$-$1.40747  &1.160 & $2.11\pm0.09$ &$-$0.96 &10.2    &$3.60   \pm0.20$ &$100.17   \pm0.55 $  &$-$0.886 &FRII   &V  V  R &  F22\\
  J0519.6+2744    &79.88762 &27.73452    &0.068 & $2.02\pm0.16$ &$-$0.54 &156.6   &$133.71 \pm0.35$ &$1697.87  \pm2.54 $  &$-$0.770 &FRI    &R  V  R &  B14\\
  J0519.6$-$4544  &79.95716 &$-$45.77886 &0.035 & $2.57\pm0.12$ &$-$1.09 &496.8   &$1106.36\pm1.79$ &$64940.10 \pm27.13$  &$-$1.474 &FRII   &R  R  R &  B02\\
  J0522.9$-$3628  &80.74160 &$-$36.45856 &0.056 & $2.45\pm0.01$ &$-$0.63 &38.9    &$2488.72\pm3.95$ &$17839.80 \pm16.31$  &$-$0.496 &FRI$-$II &V  V  R &  S06\\
  J0525.6$-$2008  &81.36683 &$-$20.18011 &0.092 & $2.22\pm0.11$ &$-$0.19 &511.8   &$112.27 \pm0.35$ &$250.24   \pm1.67 $  &0.216  &FRI    &R  V  R &  Pa21\\
  J0539.2$-$6333  &84.73453 &$-$63.54380&$-$999 & $1.96\pm0.14$ &$-$0.68 &21.6    &$9.69   \pm0.16$ &$72.23    \pm0.62 $  &$-$0.535 &FRII   &R  R  R &  \\
  J0550.5$-$3216  &87.66904 &$-$32.27124 &0.069 & $1.83\pm0.07$ &$-$0.44 &133.6   &$96.19  \pm0.50$ &$472.95   \pm1.50 $  &$-$0.295 &FRI    &R  V  R &  F76\\
  J0551.8$-$3517  &87.92625 &$-$35.25921 &0.332 & $2.10\pm0.17$ &$-$0.89 &110.3   &$28.48  \pm0.35$ &$88.06    \pm0.83 $  &0.054  &FRII   &R  V  R &  F22\\
  J0627.0+2623    &96.72875 &26.38883    &0.157 & $1.60\pm0.19$ &$-$0.57 &22.2    &$129.23 \pm0.32$ &$1413.70  \pm1.57 $  &$-$0.672 &FRI$-$II &V  V  R &  B14\\
  J0627.0$-$3529  &96.77804 &$-$35.48758 &0.054 & $1.91\pm0.03$ &$-$0.65 &214.0   &$1417.74\pm0.42$ &$4597.49  \pm3.95 $  &$-$0.057 &FRI    &R  V  R &  J09\\
  J0643.4$-$3314 &100.89416 &$-$33.24182&$-$999 & $2.44\pm0.16$ &$-$0.66 &14.8    &$58.01  \pm0.23$ &$218.08   \pm0.51 $  &$-$0.166 &FRII   &V  V  R &  \\
  J0656.3+4235    &104.04442 &42.61743   &0.059 & $1.97\pm0.11$ &$-$0.50  &54.9    &$190.05 \pm0.22$ &$940.97   \pm0.74 $  &$-$0.302 &FRI    &V  V  R &  L98\\
  J0708.9+4839    &107.28335 &48.61548   &0.019 & $1.72\pm0.14$ &$-$0.66 &443.2   &$140.44 \pm0.20$ &$715.09   \pm2.56 $  &$-$0.330 &FRI    &R  V  R &  Z12\\
  J0715.7$-$1128 &108.72867 &$-$11.56226&$-$999 & $2.61\pm0.09$ &$-$0.68 &43.9    &$9.83   \pm0.28$ &$305.80   \pm1.18 $  &$-$1.204 &FRII   &V  V  R &  \\
  J0733.0+4915    &113.24325 &49.28264   &0.668 & $2.23\pm0.19$ &$-$0.58 &16.2    &$41.20  \pm0.19$ &$122.11   \pm1.21 $  &0.160  &FRI    &V  V  R &  $^{**}$\\
  J0758.7+3746    &119.61711 &37.78661   &0.042 & $2.26\pm0.13$ &$-$0.59 &187.1   &$189.81 \pm0.29$ &$2687.06  \pm2.80 $  &$-$0.830 &FRI    &R  V  R &  A20\\
  J0828.6$-$0747  &127.22741 &$-$7.81488 &0.415 & $2.45\pm0.10$ &$-$0.75 &20.6    &$18.25  \pm0.22$ &$278.75   \pm0.81 $  &$-$0.759 &FRI$-$II &V  V  R &  F22\\
  J0829.0+1755    &127.27011 &17.90440   &0.089 & $2.15\pm0.08$ &$-$0.48 &85.8    &$132.20 \pm0.22$ &$298.00   \pm1.24 $  &0.206  &FRI$-$II &R  V  R &  A20\\
  J0840.8+1317    &130.19828 &13.20654   &0.681 & $2.48\pm0.10$ &$-$0.72 &10.0    &$825.86 \pm1.25$ &$2947.94  \pm2.71 $  &0.045  &FRII   &V  V  R &  A20\\
  J0850.4$-$2558 &132.38361 &$-$26.03034&$-$999 & $2.44\pm0.21$ &$-$0.96 &66.2    &$2.67   \pm0.27$ &$25.21    \pm0.62 $  &$-$0.652 &FRII   &V  V  R &  \\
  J0858.1+1405    &134.67269 &14.16243   &1.048 & $2.46\pm0.11$ &$-$0.84 &10.7    &$354.83 \pm0.57$ &$2438.45  \pm0.69 $  &$-$0.245 &FRII   &V  V  R &  A20\\
  J0904.0+2724    &135.88755 &27.32437   &1.723 & $2.75\pm0.07$ &$-$0.83 &17.4    &$80.71  \pm0.29$ &$207.34   \pm0.78 $  &0.427  &FRI$-$II &V  V  R &  A20\\
  J0903.5+4057    &135.92993 &40.91753   &0.890 & $2.03\pm0.19$ &$-$0.91 &95.0    &$2.96   \pm0.29$ &$9.15     \pm0.45 $  &0.176  &FRII   &L  V  R &  A20\\  
  J0912.9$-$2102  &138.25091 &$-$21.05583&0.198 & $1.82\pm0.04$ &$-$0.26 &86.7    &$166.88 \pm0.49$ &$368.61   \pm0.96 $  &0.255  &FRII   &R  V  R  & J09\\
  J0913.9+1732    &138.49092 &17.57008  &$-$999 & $2.34\pm0.20$ &$-$0.78 &4.8     &$32.88  \pm0.25$ &$101.49   \pm0.41 $  &$-$0.045 &FRII   &V  V  R  & \\
  J0920.2$-$3835  &140.05929 &$-$38.58072&0.145 & $2.02\pm0.17$ &$-$0.95 &112.6   &$91.03  \pm0.29$ &$703.63   \pm2.40 $  &$-$0.506 &FRI$-$II &R  V  R  & B14\\
  J0929.3$-$2414  &142.36767 &$-$24.27579&0.208 & $2.12\pm0.14$ &$-$0.74 &92.8    &$37.39  \pm0.29$ &$240.52   \pm1.43 $  &$-$0.394 &FRI$-$II &R  V  R  & P21\\
  J0948.0$-$3859 &147.14675 &$-$38.98312&$-$999 & $2.64\pm0.14$ &$-$0.85 &132.3   &$4.53   \pm0.30$ &$86.99    \pm1.01 $  &$-$0.985 &FRI    &R  V  R  & \\
  J0957.3$-$3851 &149.64464 &$-$38.85703&$-$999 & $2.90\pm0.15$ &$-$0.78 &21.6    &$72.39  \pm0.31$ &$322.09   \pm0.68 $  &$-$0.263 &FRII   &V  V  R  & \\
  J0958.3$-$2656  &149.60245 &$-$26.92666&0.009 & $2.13\pm0.14$ &$-$0.46 &175.4   &$179.47 \pm0.36$ &$325.31   \pm0.92 $  &0.368  &FRI    &R  V  R  & O08\\
  J1008.2$-$1000  &152.00977 &$-$9.98878 &1.688 & $3.02\pm0.15$ &$-$1.01 &14.6    &$35.19  \pm0.35$ &$627.34   \pm0.60 $  &$-$0.608 &FRII   &V  V  R  & K18\\
  J1008.0+0028    &152.04765 &0.49998    &0.097 & $2.15\pm0.07$ &$-$0.49 &48.5    &$74.97  \pm0.27$ &$472.82   \pm1.65 $  &$-$0.418 &FRI$-$II &V  V  R  & P21\\
  J1106.2$-$1048  &166.52338 &$-$10.81481&0.242 & $2.23\pm0.13$ &$-$0.67 &27.1    &$34.73  \pm0.29$ &$266.35   \pm0.77 $  &$-$0.474 &FRII   &V  V  R  & F22\\
  J1116.6+2915    &169.14424 &29.25475   &0.046 & $1.68\pm0.26$ &$-$0.58 &63.1    &$38.89  \pm0.54$ &$1891.66  \pm2.14 $  &$-$1.387 &FRII   &V  V  R  & A20\\
  J1121.3$-$0011  &170.33092 &$-$0.22134 &0.099 & $1.92\pm0.14$ &$-$0.52 &145.4   &$116.22 \pm0.33$ &$641.30   \pm1.74 $  &$-$0.347 &FRI    &R  V  R  & G23\\
  J1139.6+1149    &175.11538 &12.05206   &0.081 & $2.78\pm0.15$ &$-$0.70  &174.4   &$32.71  \pm0.27$ &$1462.91  \pm2.23 $  &$-$1.339 &FRI    &R  V  R  & A20\\
  J1144.9+1937    &176.27087 &19.60631   &0.021 & $2.01\pm0.08$ &$-$0.73 &616.9   &$360.47 \pm0.39$ &$5513.38  \pm8.11 $  &$-$0.873 &FRI    &R  V  R  & A20\\
  J1146.4$-$3327  &176.61854 &$-$33.47850&0.294 & $2.43\pm0.13$ &$-$0.66 &25.6    &$223.32 \pm0.33$ &$1591.96  \pm1.60 $  &$-$0.423 &FRII   &V  V  R  & M11\\
  J1202.4+4442    &180.53609 &44.73957   &0.297 & $2.47\pm0.18$ &$-$0.52 &178.1   &$45.86  \pm0.37$ &$104.86   \pm0.99 $  &0.256  &FRI    &L  V  R  & A20\\
  J1205.7$-$2635  &181.38838 &$-$26.56790&0.789 & $2.53\pm0.05$ &$-$0.55 &15.4    &$358.26 \pm0.66$ &$1535.06  \pm1.23 $  &$-$0.040 &FRII   &V  V  R  & W83\\
  J1216.1+0930    &184.02587 &9.48600    &0.093 & $2.05\pm0.07$ &$-$0.10  &285.6   &$92.65  \pm0.34$ &$183.94   \pm1.58 $  &0.312  &FRI    &R  V  R  & A20\\
  J1219.6+0550    &184.84674 &5.82491    &0.007 & $2.17\pm0.13$ &$-$0.47 &534.9   &$296.56 \pm0.40$ &$17724.20 \pm10.32$  &$-$1.492 &FRI    &R  V  R  & G04\\
  J1226.9+6405    &186.59378 &64.10612   &0.110 & $2.67\pm0.14$ &$-$0.18 &411.5   &$43.66  \pm0.25$ &$608.56   \pm1.72 $  &$-$0.021 &FRI    &L  V  L  & A20\\
  J1227.8+6053    &186.93893 &60.87166   &0.634 & $2.16\pm0.21$ &$-$0.79 &31.5    &$22.23  \pm0.28$ &$522.17   \pm0.42 $  &$-$0.126 &FRI$-$II &L  V  L  & A20\\
  J1230.8+1223    &187.70593 &12.39112   &0.004 & $2.06\pm0.03$ &$-$0.68 &658.7   &$2821.31\pm6.96$ &$158617.00\pm237.7$9 &$-$1.466 &FRI    &R  V  R  & B09\\
  J1230.9+3711    &187.85039 &37.18392   &0.218 & $2.43\pm0.24$ &$-$1.00  &186.9   &$16.61  \pm0.22$ &$186.42   \pm1.42 $  &$-$0.666 &FRI$-$II &L  V  R  & A20\\
  J1233.6+5027    &188.45527 &50.43965   &0.206 & $2.20\pm0.08$ &$-$0.54 &18.5    &$71.49  \pm0.24$ &$1085.35  \pm0.81 $  &$-$0.031 &FRII   &V  V  L  & A20\\
  J1236.9$-$7232  &189.29889 &$-$72.59203&0.023 & $2.43\pm0.11$ &$-$0.88 &789.2   &$308.54 \pm0.46$ &$970.18   \pm3.89 $  &$-$0.048 &FRI    &R  R  R  & L11\\
  J1256.9+2736    &194.35147 &27.49790   &0.024 & $2.68\pm0.14$ &$-$0.82 &146.1   &$3.53   \pm0.20$ &$77.66    \pm0.68 $  &$-$1.039 &FRI    &R  V  R  &  A20\\
  J1258.7+5143    &194.60580 &51.70725   &0.463 & $2.31\pm0.14$ &$-$0.56 &47.4    &$3.72   \pm0.28$ &$263.07   \pm0.40 $  &$-$0.656 &FRII   &V  V  L  & Z12\\
  J1302.7+4750    &195.70292 &47.91963   &0.140 & $2.56\pm0.26$ &$-$0.45 &158.4   &$4.31   \pm0.30$ &$204.52   \pm0.55 $  &$-$0.566 &FRI    &L  V  L  &  A20\\
  J1305.9+3858    &196.38000 &38.92250   &0.376 & $2.20\pm0.15$ &$-$0.65 &120.6   &$10.89  \pm0.24$ &$40.49    \pm0.53 $  &$-$0.049 &FRI    &L  V  R  &  A20\\
  J1306.3+1113    &196.58019 &11.22772   &0.085 & $1.87\pm0.14$ &$-$0.73 &112.4   &$93.98  \pm0.26$ &$417.11   \pm1.55 $  &$-$0.233 &FRI    &R  V  R  &  A20\\
  J1306.7$-$2148  &196.6752  &$-$21.79749&0.126 & $2.17\pm0.07$ &$-$0.48 &285.0   &$81.99  \pm0.32$ &$578.07   \pm2.30 $  &$-$0.466 &FRI    &R  V  R  & G83\\
  J1310.6+2449    &197.66051 &24.80636   &0.226 & $2.00\pm0.09$ &$-$0.70  &63.6    &$38.58  \pm0.27$ &$116.63   \pm0.66 $  &0.040  &FRII   &L  V  R  & D20\\
  J1311.8+2057    &197.93103 &20.86903   &0.724 & $2.84\pm0.76$ &$-$0.76 &25.1    &$6.95   \pm0.36$ &$157.35   \pm0.87 $  &$-$0.871 &FRII   &V  V  R  & F22\\
  J1325.5$-$4300  &201.36508 &$-$43.01911&0.001 & $2.57\pm0.02$ &$-$0.65 &901.2   &$4645.56\pm2.87$ &$246883.00\pm56.38$  &$-$1.442 &FRI    &R  R  R  & L11\\
  J1326.2+4115    &201.49559 &41.25008   &0.309 & $2.45\pm0.10$ &$-$0.46 &127.3   &$61.15  \pm0.20$ &$141.37   \pm0.80 $  &0.251  &FRI    &L  V  R  &  A20\\
  J1327.0+3154    &201.88215 &31.85755   &0.239 & $2.20\pm0.23$ &$-$0.56 &21.4    &$172.55 \pm0.72$ &$1393.10  \pm2.32 $  &$-$0.500 &FRII   &V  V  R  &  A20\\
  J1334.5+5634    &203.65622 &56.52996   &0.342 & $2.35\pm0.16$ &$-$0.55 &27.0    &$74.67  \pm0.23$ &$844.24   \pm0.49 $  &0.144  &FRII   &V  V  L  &  A20\\
  J1340.1+3857    &204.70827 &38.85298   &0.246 & $2.54\pm0.12$ &$-$0.96 &12.7    &$76.32  \pm0.26$ &$3549.63  \pm1.35 $  &$-$1.307 &FRII   &V  V  R  & B09\\
  J1341.2+3958    &205.27127 &39.99594   &0.172 & $1.79\pm0.08$ &$-$0.56 &123.5   &$31.46  \pm0.36$ &$87.24    \pm0.59 $  &0.081  &FRI    &L  V  R  &  A20\\
  J1342.7+0505    &205.68176 &5.07561    &0.136 & $2.20\pm0.09$ &$-$0.62 &83.8    &$379.75 \pm0.38$ &$1771.29  \pm2.37 $  &$-$0.245 &FRI    &R  V  R  &  A20\\
  J1344.4$-$3656 &206.09910 &$-$36.94133&$-$999 & $2.16\pm0.09$ &$-$0.70  &15.2    &$55.63  \pm0.35$ &$491.88   \pm0.94 $  &$-$0.620 &FRII   &V  V  R  & \\
  J1346.5+5330    &206.43898 &53.54785   &0.135 & $2.68\pm0.08$ &$-$0.56 &187.4   &$119.69 \pm0.44$ &$1629.20  \pm1.63 $  &$-$0.002 &FRI    &L  V  L  &  A20\\
  J1346.3$-$6026  &206.70458 &$-$60.40833&0.012 & $2.39\pm0.04$ &$-$2.14 &985.5   &$4455.95\pm2.78$ &$20844.80 \pm39.37$  &$-$0.286 &FRI$-$II &R  R  R  & W89\\
  J1347.9$-$0618  &207.02731 &$-$6.35807&$-$999 & $2.14\pm0.17$ &$-$0.77 &27.6    &$14.76  \pm0.29$ &$248.29   \pm0.95 $  &$-$0.924 &FRII   &V  V  R  & \\
  J1352.6+3133    &208.07420 &31.44628   &0.045 & $2.64\pm0.14$ &$-$0.62 &217.9   &$1994.23\pm1.74$ &$4762.99  \pm4.73 $  &0.148  &FRII   &L  V  R  &  A20\\
  J1354.8$-$1041  &208.69382 &$-$10.68407&0.330 & $2.47\pm0.04$ &$-$0.35 &29.6    &$375.03 \pm0.35$ &$709.36   \pm1.05 $  &0.424  &FRII   &V  V  R  & W83\\
  J1402.6+1600    &210.68547 &15.99907   &0.244 & $2.19\pm0.12$ &$-$0.54 &18.7    &$277.80 \pm0.46$ &$884.87   \pm0.85 $  &0.011  &FRII   &V  V  R  &  A20\\
  J1413.1$-$6519  &213.29123 &$-$65.33887&0.001 & $2.24\pm0.10$ &$-$0.63 &306.3   &$582.91 \pm0.31$ &$1335.45  \pm2.29 $  &0.164  &FRI    &R  R  R  & K04\\
  J1435.5+2021    &218.84141 &20.35496   &0.748 & $2.20\pm0.09$ &$-$0.71 &19.9    &$56.43  \pm0.28$ &$353.93   \pm0.62 $  &$-$0.253 &FRII   &V  V  R  &  A20\\
  J1443.1+5201    &220.76150 &52.02702   &0.141 & $2.11\pm0.11$ &$-$0.73 &33.3    &$207.54 \pm0.58$ &$14286.90 \pm8.54 $  &$-$0.731 &FRII   &V  V  L  &  A20\\
  J1453.0$-$1318 &223.24231 &$-$13.32303&$-$999 & $2.43\pm0.13$ &$-$0.66 &12.1    &$281.46 \pm0.37$ &$766.56   \pm0.96 $  &0.038  &FRII   &V  V  R  & \\
  J1455.4$-$3654  &223.79009 &$-$36.91874&0.094 & $2.43\pm0.11$ &$-$0.79 &234.5   &$209.22 \pm0.53$ &$1145.48  \pm2.72 $  &$-$0.345 &FRI    &R  V  R  & J09\\
  J1512.2+0202    &228.06559 &2.05471    &0.219 & $2.16\pm0.02$ &$-$0.65 &30.5    &$121.60 \pm0.34$ &$981.16   \pm2.50 $  &$-$0.505 &FRII   &R  V  R  &  A20\\
  J1516.5+0015    &229.16758 &0.25052    &0.052 & $2.55\pm0.08$ &$-$0.28 &247.3   &$744.88 \pm0.80$ &$2317.52  \pm5.66 $  &$-$0.032 &FRII   &R  V  R  &  A20\\
  J1516.8+2918    &229.17330 &29.30255   &0.130 & $1.95\pm0.15$ &$-$0.48 &82.2    &$51.56  \pm0.25$ &$129.43   \pm0.97 $  &0.138  &FRII   &L  V  R  &  A20\\
  J1518.6+0614    &229.69053 &6.23225    &0.102 & $1.80\pm0.16$ &$-$0.57 &73.8    &$111.67 \pm0.27$ &$457.98   \pm1.81 $  &$-$0.183 &FRI    &R  V  R  &  A20\\
  J1521.1+0421    &230.34393 &4.34170    &0.052 & $2.07\pm0.13$ &$-$0.23 &247.1   &$180.59 \pm0.73$ &$437.15   \pm4.50 $  &0.140  &FRI    &R  V  R  &  A20\\
  J1525.7$-$4334  &231.30062 &$-$43.68613&0.054 & $2.71\pm0.13$ &$-$0.63 &256.0   &$127.39 \pm0.32$ &$545.31   \pm1.48 $  &$-$0.223 &FRI    &R  R  R  & B14\\
  J1541.1+3451    &235.24478 &34.87312   &0.233 & $2.59\pm0.19$ &$-$0.59 &18.7    &$43.39  \pm0.44$ &$164.65   \pm0.59 $  &$-$0.099 &FRII   &V  V  R  &  A20\\
  J1556.1+2812    &239.04852 &28.19276   &0.208 & $1.88\pm0.18$ &$-$0.54 &17.6    &$54.97  \pm0.23$ &$99.14    \pm0.56 $  &0.436  &FRI    &V  V  R  &  A20\\
  J1556.6+1417    &239.19008 &14.26365   &0.350 & $2.48\pm0.18$ &$-$0.50  &94.8    &$41.08  \pm0.28$ &$130.26   \pm1.30 $  &0.042  &FRI    &R  V  R  & $^{**}$\\
  J1606.0+0011    &241.55285 &0.00755    &0.059 & $2.71\pm0.13$ &$-$0.63 &88.9    &$125.87 \pm0.54$ &$2209.39  \pm3.50 $  &$-$0.924 &FRI    &R  V  R  & B99\\
  J1606.4+1814    &241.56678 &18.24995   &0.036 & $1.89\pm0.17$ &$-$0.08 &344.1   &$183.53 \pm0.24$ &$291.73   \pm1.51 $  &0.517  &FRI    &R  V  R  &  A20\\
  J1615.6+4712    &243.9217  &47.18667   &0.198 & $2.21\pm0.06$ &$-$0.61 &86.4    &$80.91  \pm0.28$ &$568.82   \pm1.48 $  &$-$0.443 &FRI    &L  V  R  &  A20\\
  J1636.3+7128    &248.96716 &71.48162   &0.171 & $2.83\pm0.12$ &$-$0.67 &12.4    &$180.94 \pm0.26$ &$274.95   \pm0.60 $  &0.339  &FRII   &V  V  V  & A98\\
  J1643.0$-$7714  &251.06716 &$-$77.26355&0.043 & $2.51\pm0.17$ &$-$0.70  &253.5   &$324.72 \pm0.55$ &$6189.11  \pm4.57 $  &$-$0.967 &FRII   &R  R  R  & S93\\
  J1644.2+4546    &251.08321 &45.77899   &0.225 & $1.86\pm0.13$ &$-$0.41 &27.7    &$61.22  \pm0.29$ &$160.94   \pm0.67 $  &0.133  &FRII   &V  V  R  &  A20\\
  J1716.5$-$5631 &259.09019 &$-$56.42876&$-$999 & $2.75\pm0.14$ &$-$1.43 &37.6    &$7.96   \pm0.21$ &$29.50    \pm0.55 $  &$-$0.157 &FRII   &R  R  R  & \\
  J1716.3$-$3421 &259.09872 &$-$34.30417&$-$999 & $2.49\pm0.10$ &$-$0.88 &14.5    &$82.39  \pm0.42$ &$531.14   \pm0.63 $  &$-$0.461 &FRII   &V  V  R  & \\
  J1720.2+3824    &260.04306 &38.43226   &0.452 & $2.42\pm0.12$ &$-$0.65 &108.3   &$151.34 \pm0.28$ &$244.77   \pm0.85 $  &0.614  &FRI$-$II &L  V  R  &  A20\\
  J1727.9$-$0654  &261.96545 &$-$6.96904&$-$999 & $2.62\pm0.13$ &$-$0.81 &6.6     &$94.00  \pm0.28$ &$432.36   \pm0.56 $  &$-$0.281 &FRII   &V  V  R  & \\
  J1738.0+0236    &264.39465 &2.61400    &0.177 & $2.82\pm0.16$ &$-$0.57 &104.8   &$4.68   \pm0.29$ &$447.31   \pm1.57 $  &$-$1.644 &FRII   &V  V  R  & F22\\
  J1744.8+5540    &266.23587 &55.70476   &0.030 & $2.08\pm0.16$ &$-$0.41 &23.7    &$342.05 \pm0.28$ &$497.13   \pm0.79 $  &0.354  &FRI    &V  V  V  & S94\\
  J1745.0$-$1953 &266.25430 &$-$19.87146&$-$999 & $2.37\pm0.17$ &$-$0.87 &10.5    &$12.89  \pm0.35$ &$154.66   \pm0.68 $  &$-$0.766 &FRII   &V  V  R  & \\
  J1745.6+3950    &266.40730 &39.85859   &0.267 & $1.96\pm0.11$ &$-$0.71 &92.4    &$174.27 \pm0.21$ &$627.52   \pm1.41 $  &$-$0.058 &FRI    &R  V  R  & L08\\
  J1804.4+5249    &271.09484 &52.82053   &0.108 & $1.88\pm0.18$ &$-$0.39 &35.0    &$4.96   \pm0.26$ &$65.38    \pm0.87 $  &$-$1.050 &FRI    &V  V  V  & B14\\
  J1812.9+4249    &273.28384 &42.81545  &$-$999 & $2.63\pm0.14$ &$-$0.72 &29.9    &$21.91  \pm0.27$ &$72.34    \pm0.95 $  &$-$0.087 &FRII   &L  V  R  & \\
  J1824.7$-$3243  &276.23304 &$-$32.71620&0.355 & $2.35\pm0.12$ &$-$0.89 &51.0    &$107.38 \pm0.62$ &$3854.94  \pm4.65 $  &$-$1.162 &FRII   &R  V  R  & M09\\
  J1840.3$-$3037 &279.93490 &$-$30.50648&$-$999 & $2.49\pm0.13$ &$-$0.71 &54.5    &$1.73   \pm0.30$ &$574.01   \pm1.07 $  &$-$2.245 &FRII   &R  V  R  & \\
  J1843.7$-$3227 &280.79364 &$-$32.34507&$-$999 & $2.27\pm0.15$ &$-$1.10  &22.4    &$1.17   \pm0.34$ &$41.39    \pm0.54 $  &$-$1.261 &FRII   &V  V  R  & \\
  J1843.4$-$4835  &280.81091 &$-$48.60644&0.110 & $2.02\pm0.13$ &$-$0.75 &204.1   &$650.66 \pm0.48$ &$3568.16  \pm3.24 $  &$-$0.340 &FRI    &R  R  R  & J09\\
  J1845.3+5605    &281.64578 &56.15798  &$-$999 & $2.58\pm0.16$ &$-$0.85 &15.7    &$8.64   \pm0.26$ &$41.78    \pm0.72 $  &$-$0.584 &FRII   &V  V  V  & \\
  J1918.6$-$7813c&290.93080 &$-$78.19609&$-$999 & $2.62\pm0.1 $&$-$1.16 &32.6    &$25.48  \pm0.16$ &$167.21   \pm0.56 $  &$-$0.470 &FRII   &R  R  R  & \\
  J1935.0$-$0553 &293.69705 &$-$5.89465 &$-$999 & $2.64\pm0.17$ &$-$0.80  &23.1    &$1.71   \pm0.26$ &$38.14    \pm0.52 $  &$-$1.054 &FRII   &V  V  R  & \\
  J1951.2$-$0951  &297.88701 &$-$9.92623 &0.153 & $2.66\pm0.17$ &$-$0.73 &77.7    &$27.76  \pm0.29$ &$689.28   \pm1.12 $  &$-$1.053 &FRI    &R  V  R  & B14\\
  J2156.0$-$6942  &329.27491 &$-$69.68991&0.028 & $2.83\pm0.09$ &$-$0.75 &71.8    &$2349.29\pm1.89$ &$25258.90 \pm11.03$  &$-$0.705 &FRII   &R  R  R  & K22\\
  J2203.3$-$5009 &330.58617 &$-$50.11176&$-$999 & $2.60\pm0.14$ &$-$0.52 &56.7    &$2.43   \pm0.16$ &$24.52    \pm0.47 $  &$-$0.684 &FRII   &R  R  R  & \\
  J2211.2$-$1325  &332.85041 &$-$13.46936&0.392 & $2.68\pm0.05$ &$-$0.62 &12.0    &$462.68 \pm0.30$ &$1347.83  \pm0.96 $  &0.108  &FRII   &V  V  R  & P21\\
  J2211.9+0821    &333.00662 &8.32125    &0.483 & $2.51\pm0.16$ &$-$0.66 &11.4    &$566.01 \pm0.55$ &$1847.77  \pm0.61 $  &0.057  &FRII   &V  V  R  & N93\\
  J2232.3+6246    &338.09526 &62.82679  &$-$999 & $2.55\pm0.11$ &$-$0.68 &34.5    &$55.55  \pm0.58$ &$344.57   \pm2.42 $  &$-$0.716 &FRII   &V  V  V  & \\
  J2234.8$-$2610 &338.62824 &$-$26.15603&$-$999 & $2.09\pm0.16$ &$-$0.63 &11.4    &$14.65  \pm0.22$ &$81.60    \pm0.50 $  &$-$0.385 &FRII   &V  V  R  & \\
  J2245.9+1544    &341.52080 &15.74316   &0.596 & $2.02\pm0.06$ &$-$0.66 &46.6    &$6.33   \pm0.19$ &$89.01    \pm0.60 $  &$-$0.679 &FRII   &V  V  R  & P19\\
  J2302.8$-$1841  &345.76239 &$-$18.69050&0.128 & $2.27\pm0.11$ &$-$0.55 &132.5   &$234.21 \pm0.56$ &$1340.91  \pm3.37 $  &$-$0.357 &FRII   &R  V  R  & J09\\
  J2326.9$-$0201  &351.72406 &$-$2.03716 &0.188 & $2.58\pm0.10$ &$-$0.83 &101.0   &$155.71 \pm0.23$ &$2400.51  \pm2.10 $  &$-$0.824 &FRII   &V  V  R  &  A20\\
  J2329.7$-$2118  &352.41742 &$-$21.22922&0.280 & $2.31\pm0.08$ &$-$0.72 &30.3    &$139.95 \pm0.34$ &$710.42   \pm0.94 $  &$-$0.250 &FRI    &V  V  R  & J09\\
  J2330.4+1230    &352.54147 &12.47460   &0.144 & $2.27\pm0.15$ &$-$0.53 &60.3    &$122.37 \pm0.28$ &$438.58   \pm0.89 $  &$-$0.091 &FRI    &R  V  R  &  A20\\
  J2341.8$-$2917  &355.37400 &$-$29.32083&0.051 & $2.27\pm0.10$ &$-$0.38 &466.8   &$141.46 \pm0.34$ &$334.90   \pm1.68 $  &0.156  &FRI    &R  V  R  & J09\\
  J2350.9$-$1416  &357.79633 &$-$14.26644&0.127 & $1.99\pm0.15$ &$-$0.40  &131.1   &$46.11  \pm0.40$ &$103.67   \pm0.92 $  &0.220  &FRII   &R  V  R  & F22\\
  J2359.3$-$2049  &359.83139 &$-$20.79889&0.096 & $1.93\pm0.07$ &$-$0.49 &23.6    &$72.57  \pm0.30$ &$457.69   \pm0.58 $  &$-$0.418 &FRI$-$II &V  V  R  & J09\\
\enddata
\tablecomments{\small The column information are as follows: (1) 4FGL name; (2) right ascension of the counterpart (J2000, in degrees); (3) declination of the counterpart (J2000, in degrees); (4) redshift (-999 for sources without redshift information); (5) \gm-ray photon index in the 0.1$-$100 GeV energy range adopted from the 4FGL-DR4 catalog; (6) radio spectral index provided in SPECFIND; (7) largest angular size (LAS, in arcsec); (8) and (9) core and total flux densities, respectively (in mJy); (10) core dominance; (11) morphological classification; (12) the survey which was used to estimate the LAS, core and total flux densities, in respective order; V=VLASS, F=FIRST, L=LOFAR, and R=RACS; and (13) reference for redshift. They are as follows: A20: \citet[][]{2020ApJS..249....3A}; J09: \citet[][]{2009MNRAS.399..683J}; S93: \citet[][]{1993ANAS..100..395S}; Z12: \citet[][]{2012RAA....12..723Z}; J00: \citet[][]{2000ApJS..126..331J}; D20: \citet[][]{2020ApNSS.365...12D}; F22: \citet[][]{2022Univ....8..587F}; M96: \citet[][]{1996MNRAS.281..425M}; P17: \citet[][]{2017ApJ...851..135P}; B09: \citet[][]{2009ANA...495.1033B}; P21: \citet[][]{2021AJ....162..177P}; H12: \citet[][]{2012ApJS..199...26H}; L17: \citet[][]{2017ApJ...846..166L}; B14: \citet[][]{2014ApJS..210....9B}; B02: \citet[][]{2002ApJS..140..143B}; S06: \citet[][]{2006ANA...457...35S}; Pa21: \citet[][]{2021MNRAS.504.3338P}; F76: \citet[][]{1976ApJ...207L..75F}; L98: \citet[][]{1998ApJS..118..127L}; O08: \citet[][]{2008AJ....135.2424O}; K18: \citet[][]{2018ApJS..235...10K}; G23: \citet[][]{2023AJ....165..127G}; M11: \citet[][]{2011MNRAS.417.2651M}; W83: \citet[][]{1983PASAu...5....2W}; G04: \citet[][]{2004AnA...417..499G}; L11: \citet[][]{2011MNRAS.416.2840L}; G83: \citet[][]{1983MNRAS.204..691G}; W89: \citet[][]{1989AnA...223...61W}; K04: \citet[][]{2004AJ....128...16K}; B99: \citet[][]{1999MNRAS.310..223B}; A98: \citet[][]{1998ApJS..117..319A}; S93: \citet[][]{1993MNRAS.262..889S}; S94: \citet[][]{1994AnAS..103..349S}; L08: \citet[][]{2008AnA...482..771L}; M09: \citet[][]{2009AnA...495..121M}; K22: \citet[][]{2022ApJS..261....2K}; N93: \citet[][]{1993ApJ...413..453N}; P19: \citet[][]{2019ApJ...871..162P}; and, $^{**}$: this work.}
\end{deluxetable*}

\section{Summary}\label{sec6}
We have carried out a radio morphological study of a sample of \gm-ray emitting AGNs present in the 4FGL-DR4 catalog to identify potential misaligned jetted sources. We summarize our findings below.
\begin{enumerate}
\item We have identified 149 \gm-ray detected misaligned AGN, significantly increasing the known sample size. This is roughly three times the previously known objects of this class, emphasizing the effectiveness of using high-resolution and sensitive radio surveys.

\item The identified sources exhibit a variety of radio morphologies, including edge-darkened (FR I), edge-brightened (FR II), hybrid, wide-angle-tailed, bent jets, and giant radio sources. This diversity indicates that \gm-ray emission is not restricted to a specific radio morphology but can arise from a range of AGN structures.

\item We confirm that core dominance is a reliable indicator of the jet viewing angle. In our sample, most of the \gm-ray emitting misaligned AGN have low core dominance values, suggesting that their jets are not aligned with our line of sight. Among the FR I and FR II radio sources, the former appears to be more core-dominated.

\item We found several sources with hybrid morphology (e.g., FR I on one side and FR II on the other) and complex structures (e.g., wide-angle tailed and bent jets). These findings suggest that the jet interactions with the surrounding environment can lead to diverse and complex radio structures. Since the \gm-ray emission from relativistic jets is highly sensitive to the jet viewing angle, the identification of such a large number of \gm-ray emitting misaligned jetted AGN raises questions on our current understanding of high-energy emission processes operating in relativistic jets.

\item Most of the \gm-ray detected misaligned AGN in our sample have optical spectra dominated by host galaxy absorption features and/or narrow emission lines, consistent with the properties of misaligned jets. This reinforces the idea that \gm-ray emission can be detected in AGN without the relativistic beaming seen in blazars.

\item The identification of \gm-ray emitting AGN in various environments, including cluster galaxies, implies significant AGN feedback processes. These jets likely impact the intergalactic medium, influencing star formation and galaxy evolution. Deeper multiwavelength observations are needed to explore these aspects.
\end{enumerate}

\acknowledgements

We thank the journal referee for constructive criticism. VSP thanks Elizabeth Mahony for providing the optical spectrum of 4FGL J1146.4-3327 or PKS 1143$-$331 in tabular format. A.D. is thankful for the support of the Proyecto PID2021-126536OA-I00 funded by MCIN / AEI / 10.13039/501100011033.

The National Radio Astronomy Observatory is a facility of the National Science Foundation operated under cooperative agreement by Associated Universities, Inc. CIRADA is funded by a grant from the Canada Foundation for Innovation 2017 Innovation Fund (Project 35999), as well as by the Provinces of Ontario, British Columbia, Alberta, Manitoba and Quebec.

LOFAR is the Low Frequency Array designed and constructed by ASTRON. It has observing, data processing, and data storage facilities in several countries, which are owned by various parties (each with their own funding sources), and which are collectively operated by the ILT foundation under a joint scientific policy. The ILT resources have benefited from the following recent major funding sources: CNRS-INSU, Observatoire de Paris and Université d'Orléans, France; BMBF, MIWF-NRW, MPG, Germany; Science Foundation Ireland (SFI), Department of Business, Enterprise and Innovation (DBEI), Ireland; NWO, The Netherlands; The Science and Technology Facilities Council, UK; Ministry of Science and Higher Education, Poland; The Istituto Nazionale di Astrofisica (INAF), Italy.

 This paper includes archived data obtained through the CSIRO ASKAP Science Data Archive, CASDA (https://data.csiro.au). This research uses services or data provided by the Astro Data Lab, which is part of the Community Science and Data Center (CSDC) Program of NSF NOIRLab.
 
The Pan-STARRS1 Surveys (PS1) and the PS1 public science archive have been made possible through contributions by the Institute for Astronomy, the University of Hawaii, the Pan-STARRS Project Office, the Max-Planck Society and its participating institutes, the Max Planck Institute for Astronomy, Heidelberg and the Max Planck Institute for Extraterrestrial Physics, Garching, The Johns Hopkins University, Durham University, the University of Edinburgh, the Queen's University Belfast, the Harvard-Smithsonian Center for Astrophysics, the Las Cumbres Observatory Global Telescope Network Incorporated, the National Central University of Taiwan, the Space Telescope Science Institute, the National Aeronautics and Space Administration under Grant No. NNX08AR22G issued through the Planetary Science Division of the NASA Science Mission Directorate, the National Science Foundation Grant No. AST-1238877, the University of Maryland, Eotvos Lorand University (ELTE), the Los Alamos National Laboratory, and the Gordon and Betty Moore Foundation.

\appendix
\section{Multiwavelength Counterparts of Unidentified Gamma-ray Sources}\label{A:sec1}
To determine the potential counterparts of 2430 unID, we relied upon the fact that blazars and radio galaxies are the most common type of astrophysical objects among the \gm-ray sources with the known counterparts. Since jetted AGN are known to radiate across the electromagnetic spectrum, an astrophysical object detected at all accessible wavelengths is expected to be the most probable counterpart of a unID. Therefore, we first identified all X-ray detected sources lying within the 95\% uncertainty regions of the optimized \gm-ray positions of unIDs provided in the 4FGL-DR4 catalog. The parent sample of X-ray sources was prepared utilizing Chandra Source Catalog \citep[][]{2010ApJS..189...37E}, eROSITA-DE data release 1 \citep[][]{2024A&A...682A..34M}, XMM-Newton serenditipious source catalog \citep[4XMM-DR13;][]{2020A&A...641A.136W}, and live Swift-X-ray Point Source Catalog \citep[][]{2023MNRAS.518..174E}. This exercise resulted in a total 8971 X-ray sources found within the 95\% uncertainty regions of 1533 \gm-ray sources. Next, we cross-matched 8971 X-ray positions with the radio sources included in the latest VLASS/RACS (depending on the source location) catalogs to determine the radio counterpart of X-ray objects. We found 637 radio and X-ray emitting objects. Then, we cross-matched them with the Wide-field Infrared Survey Explorer \citep[WISE;][]{2010AJ....140.1868W} catalog using a 5$^{\prime\prime}$ search radius which led to the identification of 489 objects also detected in the IR band. Among them, 429 are unique sources located within the 95\% \gm-ray positional uncertainty regions of unID. There are 47 unID that have more than one X-ray-radio-IR emitting objects lying within the \gm-ray uncertainty ellipses. In such cases, we carefully checked their multi-band parameters, e.g., flux brightness, to ascertain the most promising counterpart. The radio morphologies of these 476 most probable unID counterparts were then inspected following the procedures outlined in Section~\ref{sec3}.

\section{Notes on Interesting Objects}\label{A:sec5}
We briefly discuss the interesting findings about a few double-lobed radio sources below, except those that are already well-studied as radio galaxies in the literature.

{\it 4FGL~J0001.4$-$0010 or FBQS J0001$-$0011}: This object is classified as a BL Lac in the 4FGL-DR4 catalog. However, its VLASS quick-look image reveals it to be a wide-angled-tailed (WAT) FR I source. The source has a steep radio spectrum. The SDSS optical spectrum is consistent with that of a galaxy. These observations and a low core dominance suggest this source to be a misaligned AGN. There is another galaxy located at $\sim$7.5 arcsec in the northeast direction and has a redshift ($z=0.464$) similar to FBQS J0001$-$0011 ($z=0.462$). This observation indicates the \gm-ray source to be located in a cluster environment. and the observed WAT-shaped radio morphology supports this possibility.

{\it 4FGL~J0013.6+4051 or 4C +40.01}: The VLASS image of this unclassified, lobe-dominated AGN shows a hybrid morphology. A lobe with a hotspot, similar to FR IIs, is identified in the southeast direction, whereas FR I radio morphology is seen on the opposite side. Both the VLASS and LOFAR images show diffuse emission surrounding the object, reminiscent of a fat double such as 3CR~310 \citep[e.g.][]{2012ApJ...749...19K}. Its overall radio spectrum is steep. The optical spectrum of the source is red and shows galaxy absorption features and only narrow emission lines \citep[][]{1993ANAS..100..395S}. This object was recently proposed as a misaligned AGN by \citet[][]{2022Univ....8..587F}.

{\it 4FGL~J0018.8+2611 or 4C +25.01}: This object is a marginally core-dominated broad line quasar. The VLASS image resolved the source into a misaligned triple structure with two edge-brightened lobes and a bright core, which has been referred to as a dog-leg structure \citep[e.g.][and references therein]{2014ApJS..212...19F}. The high-resolution image by \citet[][]{2014ApJS..212...19F} shows two knots of emission towards the south, which could be part of a jet. The misaligned morphological features are also seen in the LOFAR and RACS images. This source was also studied by \citet[][]{1984AJ.....89.1658G} who explained the observed curved shape due to interaction of the quasar with the intergalactic medium and/or possible projection effects \citep[see also,][]{1982JApA....3..465K}. The radio spectrum of the source is steep.

{\it 4FGL~J0309.4$-$4000 or PKS 0307$-$402}: This \gm-ray source is classified as a BL Lac in the 4FGL-DR4 catalog. Its RACS image shows a complex morphology. The radio emission observed to the north of the core is likely to be associated with a different optical object ($\alpha=47^{\circ}.3045020$, $\delta=-40^{\circ}.0194409$). A lobe-like feature seen in the southeastern direction is also found to be positionally consistent with an optical source ($\alpha=47^{\circ}.3301066$, $\delta=-40^{\circ}.0476778$). Excluding these two features, the radio morphology of the source appears like a C-shaped/WAT structure. Such bent radio jets are often found in FR I radio galaxies which are located in dense cluster environments. The optical spectrum of PKS 0307$-$402 shows host galaxy absorption features similar to that typically seen in a LERG \citep[][]{2021AJ....162..177P}. Its overall radio spectrum is steep and the derived core dominance indicates the source to be a lobe-dominated AGN.

{\it 4FGL~J0550.5$-$3216 or PKS 0548$-$322}: This source is classified as a BL Lac in the 4FGL-DR4 catalog. Its optical spectrum consists of a blue non-thermal continuum along with the host galaxy absorption features \citep[][]{1976ApJ...207L..75F}. \citet[][]{1984Natur.308..617A} reported the detection of a head-tail morphology which was well resolved in the VLASS image analyzed in this work. The low-frequency observations taken with RACS also reveal this object's WAT nature and much larger diffuse emission towards the southern direction. The computed core dominance is low.

{\it 4FGL~J0904.0+2724 or B2~0900+27}: This distant \gm-ray source exhibits broad emission lines in its optical spectrum, leading to its classification as an FSRQ. Its radio spectrum is steep. The VLASS image of this object reveals a double-lobed structure with the hotspot towards the southeastern direction being clearly visible. There is a suggestion of a jet towards the northeast, which is also seen in the Very Long Baseline Array (VLBA) observations at 7.6 GHz \citep[][]{2021AJ....161...14P}. It is moderately core-dominated and is likely to be misaligned.

{\it 4FGL~J0920.2$-$3835 or MRC 0918$-$383}: This object is reported as a BCU in the 4FGL catalog. Its RACS image shows a hybrid morphology. A bright lobe and jet emission are observed towards the northwest of the core. This feature is also resolved in the VLASS data. On the other hand, a much larger diffuse emission, similar to FR Is, is identified in the southeastern direction. This feature is not seen in the VLASS data. The northwestern jet was also observed in the Molonglo Observatory Synthesis Telescope data taken at 843 MHz with 44$^{\prime\prime}$ resolution \citep[][]{1992ApJS...80..137J}. The estimated core dominance suggests MRC 0918$-$383 to be a lobe-dominated AGN. The overall radio spectrum is steep. There is no published optical spectroscopic information about this object.

{\it 4FGL~J0948.0$-$3859}: The object VLASS1QLCIR J094835.22$-$385859.2, also known as PMN J0948$-$3859, is a possible counterpart of this unidentified \gm-ray source. Its VLASS and RACS images show interesting radio structures. The inner jet is identified south of the core, whereas a diffuse, FR~I type structure is observed in the northern direction, which is further bent towards the southwest. The overall morphology is consistent with an FR I type radio galaxy. Its radio spectrum is steep and the estimated core dominance suggests it to be a lobe-dominated AGN.

{\it 4FGL~J1008.0+0028 or PKS 1005+007}: This \gm-ray source has been classified as a BL Lac in the 4FGL-DR4 catalog. Its VLASS image reveals a lobe-dominated, hybrid radio morphology. A lobe with a hotspot, similar to FR IIs, was seen in the north-eastern direction, whereas there is a larger diffuse, plume-shaped FR I radio emission on the western side. The FIRST data shows a similar morphology along with larger scale diffuse low surface brightness radio emission perpendicular to the jet direction. Indeed, this source has been proposed as an `X-shaped' radio galaxy \citep[][]{2018ApJ...852...47R}. The optical spectrum of PKS 1005+007 appears to be dominated by the galaxy absorption features, and the host galaxy is resolved in the PanSTARRS $i$-band image.

{\it 4FGL~J1121.3$-$0011 or MGC 0019706}: This \gm-ray source is classified as a BCU in the 4FGL-DR4 catalog. Its FIRST image shows a curved FR I morphology similar to WAT radio galaxies, indicating it to lie in a cluster environment \citep[see also,][]{2004MNRAS.348..866E,2014MNRAS.438..796S}. The RACS data also reveals a similar radio morphology. The source was earlier imaged by \citet[][]{1991AJ....102.1960P} in a study of dumbbell galaxies, who reported a similar structure associating the source with the brighter north-western component of the dumbbell system. The optical spectrum of the source is similar to that seen in LERGs \citep[][]{2023AJ....165..127G}. It has been reported as a misaligned AGN by \citet[][]{2022Univ....8..587F}. The derived core dominance suggests this object to be a lobe-dominated AGN and its overall radio spectrum is steep.

{\it 4FGL~J1230.9+3711 or 1RXS J123124.8+371117}: This object has been reported as a narrow-line radio galaxy with hybrid FR I/II morphology \citep[][]{2011MNRAS.415.1013K,2013ApJ...765...62S}. The LOFAR and RACS images show a lobe and a hotspot west to the core and a diffuse low surface brightness emission eastwards of the core, and thus consistent with this finding. The SDSS optical spectrum is similar to the LERG-type AGN. The overall radio spectrum of the source is steep. The computed core dominance suggests it to be a lobe-dominated AGN. It is also reported as a misaligned AGN by \citet[][]{2022Univ....8..587F}.

{\it 4FGL~J1233.6+5027 or TXS 1231+507}: This object is classified as a BL Lac in the 4FGL-DR4 catalog. The VLASS image shows edge-brightened lobes similar to FRII sources, while the lower-resolution images, especially the low-frequency LOFAR image, show plume-shaped diffuse emission on both sides. The plume on the north-western side is more extended and bends away from the axis of the source toward the north.
The LOFAR observation of TXS 1231+507 has also been studied by \citet[][]{2022MNRAS.514.2122P}, who proposed the jet reorientation as a possible explanation for the observed feature. The optical spectrum of this object is galaxy-dominated, and the radio spectrum is steep. This object has been proposed as a misaligned AGN by \citet[][]{2022Univ....8..587F}.

{\it 4FGL~J1256.9+2736}: The source VLASS1QLCIR J125724.39+272956.4, also known as NGC 4839, is the possible counterpart of this unidentified Fermi-LAT detected source. This object is reported to be a BCG in the NGC 4839 group, which is merging with the Coma cluster \citep[cf.][]{2023ApJ...944...51O}. Its VLASS image shows diffuse, bipolar radio emission similar to FR I radio galaxies. The source is also resolved in the FIRST and LOFAR datasets. The latter shows a C-shaped radio structure along with a low-surface brightness emission towards the southwestern direction. The SDSS optical spectrum comprises of absorption features arising from the host galaxy stellar population. It is a lobe-dominated AGN and has a steep radio spectrum.

{\it 4FGL~J1326.2+4115 or IVS B1323+415}: This source is classified as a BL Lac in the 4FGL-DR4 catalog. Its LOFAR data reveals curved jets hinting a C-shaped FR~I structure reminiscent of a WAT radio galaxy. Similar results are obtained by examining the FIRST and RACS datasets. The SDSS optical spectrum of this source is similar to that of a LERG or a narrow-line radio galaxy (NLRG). It is a moderately core-dominated AGN with a flat radio spectrum.

{\it 4FGL~J1327.0+3154 or 4C +32.44B}: The optical spectrum of this \gm-ray emitting BCU shows bright narrow emission lines on top of a red continuum and host galaxy absorption features. The VLASS image of this object shows it to be lobe-dominated, with an overall structure reminiscent of a WAT source. There is a prominent hotspot towards the south, with low surface brightness diffuse emission appearing to surround the source. Similar features were also seen in the FIRST and LOFAR data. Its radio spectrum is steep. This object has been reported as the brightest cluster galaxy (BCG) by \citet[][]{2011ApJ...736...21S}, consistent with its bent morphology being due to the cluster environment.

{\it 4FGL~J1340.1+3857 or 3C 288}: This \gm-ray source is classified as a BCU in the 4FGL-DR4 catalog. It has shown edge-darkened morphology in VLA observations, though its 1.5 GHz radio power ($2.5\times10^{33}$ erg cm$^{-2}$ s$^{-1}$ Hz$^{-1}$) indicates it to be an FR II radio source. \citet[][]{1989AJ.....97..674B} reported the identification of a jet and a counterjet near the radio core and extended radio wings connected with the radio lobes. Their 0.6-arcsec resolution image shows edge brightening in the inner lobes, which merge with the wings. A WAT-shaped radio morphology suggests it may be associated with a cluster of galaxies. The optical spectrum of this object is typical of a LERG \citep[][]{2009ANA...495.1033B}. It is a lobe-dominated AGN with a steep radio spectrum.

{\it 4FGL~J1341.2+3958 or RBS 1302}: This \gm-ray detected object is classified as a BL Lac in the 4FGL-DR4 catalog. It exhibits a complex morphology in the LOFAR data. A diffuse FR I jet is seen in the northeastern direction, which is connected with a large extended low-surface brightness radio emission almost throughout its length. The source is partially resolved in the FIRST and RACS datasets showing similar radio structures. It is consistent with a WAT whose tails appear superposed. The SDSS optical spectrum shows host galaxy absorption features. The estimated core dominance hints that RBS 1302 is a lobe-dominated AGN. The overall radio spectrum is steep. This object is identified as a low-power radio galaxy by \citet[][]{2018AJ....155..188L}. 

{\it 4FGL~J1342.7+0505 or 4C +05.57}: This lobe-dominated source is classified as a BL Lac in the 4FGL-DR4 catalog. It has also been identified as a narrow-line radio galaxy \citep[][]{2013ApJ...765...62S}. The VLASS image of this object shows a complex radio morphology. There is a jet spewing southwards of the radio core with a lobe/knot positionally consistent with an optical object located at $\sim$5 arcsec. At mas scale, the direction of the jet probed by VLBA is the same. This observation indicates that the secondary object could be the optical jet rather than a separate astrophysical object. Furthermore, a `Z'-shaped FR I radio morphology is seen on the large scale. The SDSS optical spectrum of 4C +05.57 is dominated by the strong narrow emission lines and host galaxy absorption features. This object has been proposed as a misaligned AGN by \citet[][]{2022Univ....8..587F}.

{\it 4FGL~J1346.5+5330 or RBS 1310}: This unclassified Fermi-LAT detected AGN has revealed a complex radio structure in the LOFAR cutout image. Closer to the core, bipolar radio jets are seen in the east-west directions. However, the large-scale jet appears to be oriented in northeastern and southwestern directions. The inner structure is also resolved in the VLASS and FIRST datasets. This `S'-shaped morphology has also been reported by \citet[][]{2019RAA....19..127L}. \citet[][]{2022MNRAS.514.2122P} explained the observed morphology due to possible jet reorientation. The source exhibits a steep radio spectrum and the computed core dominance indicates it to be a lobe-dominated AGN. Its SDSS optical spectrum has revealed broad emission lines along with host galaxy absorption features.

{\it 4FGL~J1402.6+1600 or 4C +16.39}: This object is classified as a BL Lac in the 4FGL-DR4 catalog. However, its optical spectrum shows a blue continuum with narrow emission lines and host galaxy absorption features. The radio structure of the source revealed by VLASS consists of two bright lobes with hotspots, a core, and a jet toward the eastern direction of the core. The calculated core dominance suggests the source to be marginally lobe-dominated, and the overall radio spectrum is steep. The FIRST data exhibits similar features along with larger low surface brightness emission surrounding the object. \citet[][]{1985ApJ...294..158A} analyzed the VLA data of 4C +16.39 and reported the identification of triple radio structure, i.e., a core and double lobes, with complex morphology. Higher-resolution observations show the jet to be curved with the outer prominent knot having a magnetic field structure suggesting interaction with the external environment \citep[][]{1987MNRAS.228..203S}.

{\it 4FGL~J1413.1$-$6519 or Circinus galaxy}: This source is the nearest Seyfert 2 galaxy and has been extensively studied. Its RACS image shows an X-shaped radio structure. Overplotting the RACS contours on the Dark Energy Camera image, the brighter emission, likely to be produced due to an edge-brightened jet, appears perpendicular to the galaxy disk. On the other hand, the fainter radio emission, which is aligned to the galaxy disk, is possibly originated from the starburst activities \citep[e.g.,][]{1998MNRAS.297.1202E}. Its overall radio spectrum is steep. 

{\it 4FGL~J1615.6+4712 or TXS 1614+473}: This source is classified as an FSRQ in the 4FGL-DR4 catalog. However, its SDSS optical spectrum is found to be similar to a LERG with no visible broad emission lines. The FIRST image of this source shows a complex radio morphology, suggestive of a WAT structure where the tails of emission appear merged. Moreover, the tail of extended radio emission in the northeastern direction is also visible in the LOFAR data. \citet[][]{2017MNRAS.466.4346M} reported this source to be an FR I radio galaxy based on the FIRST data. It is reported as a misaligned AGN by \citet[][]{2022Univ....8..587F}. It is a lobe-dominated AGN with a steep radio spectrum.

{\it 4FGL~J1720.2+3824 or SDSS J172010.33+382556.1}: This \gm-ray source is reported as a BCU in the 4FGL-DR4 catalog. It exhibits a hybrid morphology in the LOFAR image. An FR II jet, including a radio knot and a lobe, is identified in the southwestern direction. A diffuse FR I lobe is observed in the northwestern direction, giving an impression of an overall C-shaped radio structure. The source is partially resolved in the RACS data, whereas the FIRST image shows only the southwestern jet. It is a core-dominated AGN and has a steep radio spectrum. The SDSS optical spectrum is similar to Type 2 Seyfert galaxies. \citet[][]{2022Univ....8..587F} have suggested this object to be a misaligned AGN.

{\it 4FGL~J1745.6+3950 or RX J1745.5+3951}: Originally classified as a BL Lac, this source is also proposed as a BCG \citep[][]{2008AnA...482..771L}. Its optical spectrum exhibits strong host galaxy absorption features. The LOFAR image of this source reveals a WAT-shaped radio morphology with eastern and western jets bent southwards. This object is moderately lobe-dominated and has a steep radio spectrum.

{\it 4FGL~J1840.3$-$3037 or PKS 1836$-$305}: This Fermi-LAT detected BCU exhibits a triple radio structure in the VLASS data with a weak core. A radio-emitting knot is also resolved towards the south-eastern side. The RACS image shows the double-lobed structure with an FR~II morphology and prominent wings of emission reminiscent of an X-shaped source. The estimated core dominance parameter indicates the source to be a lobe-dominated AGN. The overall radio spectrum is steep.

{\it 4FGL~J1951.2$-$0951 or PKS 1948$-$10}: This object is classified as a BCU in the 4FGL-DR4 catalog due to the lack of optical spectroscopic information. It exhibits an inverted `C'-shaped FR~I morphology in the VLASS data indicating it to be a possible WAT radio galaxy. This source is a lobe-dominated AGN with a steep radio spectrum.

{\it 4FGL~J2211.9+0821 or PKS 2209+08}: This well-studied quasar is known to exhibit kpc-scale optical and X-ray jets \citep[e.g.,][]{2017ApJ...849...95B}. Its VLASS image exhibits a double-lobed FR II morphology. It is marginally core-dominated, and the overall radio spectrum is steep. \citet[][]{1992A&A...261...25H} reported the VLA observation of the source, which resolved it into a prominent core and a bright jet, including knots and hotspots, extending southwards. A counter-lobe with a hotspot was also seen.

{\it 4FGL~J2359.3$-$2049 or TXS 2356$-$210}: This \gm-ray detected BL Lac object was earlier studied by \citet[][]{1998ApJS..118..275K} who reported the identification of a triple radio structure. The VLASS image of this source exhibits a bright FR II lobe in the northeastern direction, while a higher resolution image is required to clarify the structure on the southwestern side. It is a lobe-dominated object whose measured radio spectral index lies at the boundary of the flat/steep spectrum definition.

\subsection{Missing Gamma-ray Emitting Radio Galaxies}\label{app:missing}
The 12 \gm-ray emitting radio galaxies present in the 4FGL-DR4 catalog are missing from our sample of double-lobed radio sources. For most of them, the primary reason is the non-detection of the bi-polar extended radio emission in the considered survey data and/or their observed properties, such as $C_{\rm D}$ and radio spectral index, not qualifying the adopted criteria. We discuss them below.

{\it 4FGL J0312.9+4119 or B3 0309+411B}: This object was first reported as a giant radio galaxy using 327 MHz observations taken with Westerbork Synthesis Radio Telescope \citep[][]{1989A&A...226L..13D}. They identified rapid flux variability at cm wavelengths and a core-dominated emission indicating a small viewing angle of the jet axis. The VLASS quick-look image of this object is affected by artifacts, possibly due to bright core emission. The RACS data taken at $\sim$1.4 GHz reveals a core-jet structure. This source is not present in the footprints of other surveys considered in this work.

{\it 4FGL J0319.8+4130 or NGC 1275}: It is an asymmetrical FR I radio source located in the Perseus cluster \citep[][]{1990MNRAS.246..477P}. Its VLASS quick-look images are affected by artifacts, possibly due to bright core emission. The RACS 1.4 GHz data shows a compact radio emission with no obvious extended structure. The source is not present in the footprints of other surveys considered in this work.

{\it 4FGL~J0433.0+0522 or 3C 120}: This object has been reported as a broad line FR I radio galaxy \citep[e.g., ][]{1989MNRAS.238..357O}. It has a high core dominance ($C_{\rm D}=0.691$), and the overall radio spectrum is flat. Its optical spectrum is also similar to Type 1 broad line quasars. Since none of the observables support its misaligned nature, this object has not appeared in our sample.
 
{\it 4FGL J0931.9+6737 or NGC 2892}: The \gm-ray source 4FGL~J0931.9+6737 has been associated with the radio galaxy NGC 2892. However, the coordinates of the counterpart given in the 4FGL-DR4 catalog refer to a BL Lac object, which is about 5$^{\prime}$ away from NGC 2892. We found that NGC 2892 lies outside of the 95\% uncertainty region of the \gm-ray position reported in the 4FGL-DR4 catalog, hence is not the correct counterpart of 4FGL~J0931.9+6737. \citet[][]{2022Univ....8..587F} also arrived at the same conclusion about this \gm-ray source.

{\it 4FGL J1149.0+5924 or NGC 3894}: This source has been identified as a compact symmetric object hosting twin parsec-scale jets \citep[][]{2024ApJ...961..240K}. Its VLASS cutout images are affected by artifacts, and it is barely resolved in the LOFAR Data. It is extremely core-dominated ($C_{\rm D}=1.90$) and has an inverted flat radio spectrum ($\alpha=0.16$). Therefore, this source did not pass the selection criteria adopted in this work to identify misaligned AGN.

{\it 4FGL J1449.5+2746 or B2 1447+27}: This source exhibited a compact core morphology in VLASS, FIRST, and RACS images. We could not find any published radio image of this object where a double-lobed morphology was identified. Moreover, \citet[][]{1986A&AS...65..111D} reported B2~1447+27 to consist of a single component with an angular size of 3.6$\times$1.4 arcsec$^2$. Therefore, it has not entered our sample.

{\it 4FGL J1530.3+2709 or LEDA 55267} and {\it 4FGL J1628.8+2529 or LEDA 58287}: These objects are identified as FR 0 radio galaxies which are objects having optical properties similar to FR~Is but lack extended radio jets \citep[][]{2018A&A...609A...1B,2021ApJ...918L..39P,2023ApJ...957...73P}.

{\it 4FGL J1724.2$-$6501 or NGC 6328}: The radio properties of this object were studied by \citet[][]{1997AJ....113.2025T} who reported it to be a GHz peaked-spectrum object. The mas scale radio observations indicated this source to be a newly born compact symmetric object \citep[][]{1997AJ....113.2025T}. NGC 6328 has a compact morphology in the RACS observation, and no extended emission was observed.

{\it 4FGL J2227.9-3031 or PKS 2225$-$308}: \citet[][]{1992ApJS...80..137J} reported the detection of a double-source using Molonglo Observatory Synthesis Telescope at 843 MHz and 44 arcsec resolution. However, the RACS image of this source revealed a compact radio object. There is a nearby galaxy (LEDA 192291) which exhibits a head-tail morphology and was also reported by \citet[][]{1989MNRAS.236..737E}. The VLASS data showed a slightly resolved radio jet in the north-west direction.

{\it 4FGL J1630.6+8234 or NGC 6251}: This object is a giant radio galaxy identified using MHz frequency observations \citep[cf.][]{1977MNRAS.181..465W}. Its VLASS quick-look image is dominated by the bright, compact core with no clear extended radio structure. This source does not lie in the footprints of other considered surveys.

{\it 4FGL J2333.9$-$2343 or PKS 2331$-$240}: This is a giant radio galaxy with a blazar nucleus, and thus, it is probably the best example of jet bending \citep[e.g.,][]{2017A&A...603A.131H}. Its RACS cutout image has revealed a bright, compact core and two lobes with the bridge emission remaining undetected \citep[see also,][]{2020MNRAS.494..902B}. It has a high core dominance ($C_{\rm D}=1.016$) and has an inverted and flat radio spectrum ($\alpha=0.13$). Therefore, this source did not pass the selection criteria adopted in this work to identify misaligned AGN.

\bibliographystyle{aasjournal}
\bibliography{Master}

\end{document}